\documentclass[iop]{emulateapj}
\usepackage{graphicx}
\usepackage{graphics}
\usepackage{amsmath}
\usepackage{multirow}
\usepackage{color}
\usepackage{xcolor}

\usepackage{float}

\def\kms{{\rm\,km\,s^{-1}}}
\def\kmskpc{{\rm\,km\, \,s^{-1} \, {kpc}^{-1}}}

\def\Myr{{\rm\,Myr}}
\def\Gyr{{\rm\,Gyr}}
\def\deg{{^\circ}}

\def\kpc{{\rm\,kpc}}

\def\mathnew{\mathsurround=0pt}   
\def\simov#1#2{\lower .5pt\vbox{\baselineskip0pt  
    \lineskip-.5pt\ialign{$\mathnew#1\hfil##\hfil$\crcr#2\crcr\sim\crcr}}}

\def\'#1{\ifx#1i{\accent"13\i}\else{\accent"13#1}\fi}    
  
\def\et{et~al. }     
 
\begin{document}

\shorttitle{Contribution of Spiral Arms to the Thick Disk}  
\shortauthors{Martinez-Medina et al. 2015}

\title{The Contribution of Spiral Arms to the Thick Disk along the Hubble Sequence}

\author{L. A. Martinez-Medina $^{1}$, B. Pichardo$^{2}$, A. P\'erez-Villegas$^{3}$ \& E. Moreno$^{2}$}
\affil{$^{1}$ Departamento de F\'isica, Centro de Investigaci\'on y de Estudios
Avanzados del IPN, A.P. 14-740, 07000 M\'exico D.F., M\'exico; lmedina@fis.cinvestav.mx}
\affil{$^{2}$Instituto de Astronom\'ia, Universidad Nacional
Aut\'onoma de M\'exico, A.P. 70--264, 04510, M\'exico D.F., M\'exico; barbara@astro.unam.mx}
\affil{$^{3}$ Centro de Radioastronom\'ia y Astrof\'isica, Universidad Nacional
Aut\'onoma de M\'exico, Apartado Postal 3--72, 58090 Morelia, Michoac\'an, M\'exico; mperez@astro.unam.mx}


\begin{abstract}
The first mechanism invoked to explain the existence of the thick disk
in the Milky Way Galaxy, were the spiral arms. Up-to-date work summon
several other possibilities that together seem to better explain this
component of our Galaxy. All these processes must affect differently
in distinct types of galaxies, but the contribution of each one has
not been straightforward to quantify. In this work, we present a first
comprehensive study of the effect of the spiral arms in the formation
of thick disks, as going from early to late type disk galaxies, in an
attempt to characterize and quantify this specific mechanism in
galactic potentials. To this purpose, we perform numerical simulations
of test particles in a three-dimensional spiral galaxy potential of
normal spiral galaxies (from early to late types). By varying the
parameters of the spiral arms we found that the vertical heating of
the stellar disk becomes very important in some cases, and strongly
depends on the galaxy morphology, pitch angle, arms mass and its
pattern speed. The later the galaxy type, the larger is the effect on
the disk heating. This study shows that the physical mechanism causing 
the vertical heating is different from simple resonant excitation. The 
spiral pattern induce chaotic behavior not linked necessarily to 
resonances but to direct scattering of disk stars, which leads to an 
increase of the velocity dispersion. We applied this study to the specific
example of the Milky Way Galaxy, for which we have also added an 
experiment that includes the Galactic bar. From this study we deduce
that the effect of spiral arms of a Milky-Way-like potential, on the
dynamical vertical heating of the disk is negligible, unlike later
galactic potentials for disks.

\end{abstract}

 \keywords{galaxies: evolution --- galaxies: kinematics and dynamics --- galaxies: spiral --- galaxies: structure}

\section{Introduction} \label{sec:intro}
Simulations of galaxy formation are coming to point where detailed processes 
of galaxies have never  been explored before in detail, such as random and 
rotational velocities can be better studied and understood
\citep{Scannapieco2011}. Details on disk potentials can be probed
and compared with observations, and we are now able to shed some light
on evolution of galaxies starting now on small scale stellar motions.

Dynamical heating of the Milky Way disk has now been known for over 60
years, mainly through observations in the solar vicinity. From those
observations we learnt that stellar random motions correlate nicely
with their ages known as the age-$\sigma$ relation \citep{Wielen1977, 
Binney2000}. In particular, in the case of the Milky Way disk, it is
known that the radial velocity dispersion is twice as much as the
vertical dispersion and that the radial scale length of the thick disk 
is much shorter than that of the thin disk \citep{Bensby2011}.

Recent studies also show that many, if not all, edge-on spiral
galaxies appear to host dual disk systems
\citep{Gerssen2012,vanderKruit1981}, a younger, dynamically colder and
thinner component: the thin disk and at least one older component,
(mainly) stellar, dynamically hotter and thicker component: the thick
disk \citep{Yoachim2006, Yoachim2008, Comeron2011}.

It is still not straightforward to dilucidate the mechanism
responsible for the vertical heating of the disk from observations,
specially since more likely it might be rather a combination of
several possibilities. Among the mechanisms proposed there are some
external to the disk such as hits by satellite galaxies or minor
mergers \citep{Huang1997, Velazquez1999, Benson2004, Font2001, 
Quinn1993, Villalobos2008, DiMatteo2011}; scattering by dark halo
objects or globular clusters \citep{VandePutte2009, Hanninen2002}. And
some of internal origin, such as, dynamical heating by direct
encounters with giant molecular clouds \citep{Carlberg1987, 
Villumsen1983, Lacey1985, Lacey1984, Spitzer1951, Inoue2014};
heating by encounters with the potential produced by long-lasting
spiral arms \citep{Faure2014} or irregular and transient spiral
structure \citep{Minchev2006, Fuchs2001, Jenkins1990, Barbanis1967};
perturbations from stellar bars \citep{Saha2010}; dissolution of
young stellar clusters \citep{Kroupa2002}; or during an intense 
star formation phase in a period of intense accretion very early in 
the history of the Galaxy \citep{Snaith2014}.

Because of the nature of these theories, the effects are dependent on
galaxy morphology, particularly the intrinsic mechanisms such as bars
and spiral arms. Therefore, to deeply understand secular evolution of
disk galaxies, it is critical to study dynamical heating in a good
sample of different disk galaxy types. Finally, from the point of view
of observations, the radial heating agent seems to vary exactly as
expected if the agent were the spiral arms, which provides a good
chance that the spiral structure has at least an important role as a
heating mechanism in the plane of galactic disks
\citep{Gerssen2012}. The importance and influence of spiral arms, and
even their very same nature is still under debate, there is no
straightforward observational prove yet of their effect on stars,
however it is nowadays considered to have a key role on large-scale
galactic dynamics \citep{Sellwood2013}, for a review,
\citep{Roskar2012, Minchev2012, Lepine2011, Quillen2011,
  Antoja2009}. One plausible observational example are the stellar
features seen in the velocity space, known as ``moving groups''
\citep{Proctor1869, Eggen1959, Eggen1977, Eggen1990,
  Eggen1996a,Eggen1996b, Wilson1923, Roman1949, Soderblom1993,
  Majewski1994, Majewski1996}; these structures might become the first
clear, undirect though, evidence of the effect of the spiral arms
\citep{Chereul1998, Chereul1999, Dehnen1998, Famaey2005, Famaey2008,
  Antoja2012, Pompeia2011}. Of course, the spiral arms are not the
only nor the prefered mechanism to explain moving groups, the galactic
bar is other possibility.

In this work we focus on the very first proposal to explain the
vertical heating \citep{Wielen1977}, disregarded at some point in the
history because of their negligible effect on the Milky Way disk: the
dynamical heating by effect of the spiral arms. We attempt to isolate
and quantify the contribution of the spiral arms to the disk heating
of galaxies. We performed numerical simulations of test particles in a
three-dimensional galactic potential that models spiral arms
\citep{PMME03}, adjusted to simulate spiral galaxies, from early to
late types \citep{Perez2012, Perez2013}. We produced a set-up with
relaxed initial conditions for a stellar disk. Finally we calculated
the effect on the vertical heating of the stellar disk produced by the
nonaxisymmetric large scale structures. We have included a preliminary
study on a self-gravitating potential, known as PERLAS model
\citep{PMME03,Pichardo2004} for the Milky Way Galaxy, that includes
the spiral arms and the galactic bar.

This paper is organized as follows. The galactic models, initial
conditions and methodology are described in Section \ref{model}. 
The role of each one of the parameters of the model is studied 
with detail in Section \ref{results}, where we present
calculations of dispersion velocity, the velocity ellipsoid, 
time evolution for spiral galaxies from early to late types, and 
an application to the Milky Way Galaxy. Finally, in Section 
\ref{conclusions}, we present our conclusions.

\section{Methodology and Numerical Implementation}\label{model}
The effect of the spiral arms over the stellar disk has been studied
profusely in either N-body simulations \citep{Sellwood2002,
  Roskar2012, Roskar2013, Kawata2014} or with spiral patterns treated
as perturbations to the axisymmetric background and modelling it as a
density wave \citep{DeSimone2004, Minchev2006, Faure2014}.

In N-body simulations, although self-consistent, it is not plausible
to adjust a given specific galaxy, or to isolate the effect for
example, of the arms or establish in detail the role that each one of
the parameters that characterize the spiral pattern play over the
vertical heating of the disk. On the other hand, when treated as
steady spirals exerted as a perturbation to the axisymmetric
potential, according to the hypothesis of \citet{Lin1969}, the heating
is minimum and linked only to the resonant regions of the spirals
\citep{Lynden1972, DeSimone2004}, being more efficient in the radial
direction \citep{Sellwood1984, Minchev2006}.

Here we use spiral arms that although steady, are very different in
nature to the typically and widely employed in literature. The
gravitational potential due to the spiral pattern is not a simple
perturbation but is rather based on a mass density distribution. With
this model of spiral arms our studies on the vertical heating of the
disk contrast in general with the density wave approach (except of
course, for really small spiral arm masses and pitch angles).

For the orbital study we employed then three-dimensional galactic
potentials to model normal spiral galaxies (Sa, Sb and Sc). The motion
equations are solved in the non-inertial reference system of the
spiral arms and in Cartesian coordinates ($x',y',z'$). The orbits are
integrated for 5 Gyr with a Bulirsh-Stoer algorithm (Press et
al. 1992), with a conservation of the Jacobi constant approximately up
to $10^{-12}$. The disk heating is computed through the measure of the
velocity dispersion at different times.

\subsection{Models for Normal Spiral Galaxies}\label{normal_spiral}

The models include an axisymmetric component (bulge, massive halo and
disk), as the background potential, formed by a Miyamoto \& Nagai
(1975) disk and bulge, and a massive halo \citep{Allen1991}. The
parameters used to model normal spiral galaxies (Sa, Sb and Sc) are
presented in Table \ref{tab:parameters} (compiled by \citet{Perez2013}).

Superposed to the axisymmetric components, for the spiral arms
potential, we employed a bisymmetric self-gravitating
three-dimensional potential, based on a density distribution, called
PERLAS model \citep{PMME03}. This potential consists of individual
inhomogeneous oblate spheroids superposed along a logarithmic spiral
locus \citep{RHV79}. Each spheroid has a similar mass distribution,
i.e., surfaces of equal density are concentric spheroids of constant
semiaxis ratio. The model considers a linear fall in density within
each spheroid. The minor and major semiaxes of each oblate spheroid
are 0.5 and 1.0 $\kpc$, respectively (this gives a width of the spiral
arms of 2 $\kpc$ and height of 0.5 $\kpc$ from the disk plane) and the
separation among the spheroid centers along the spiral locus is 0.5
$\kpc$. The superposition of the spheroids begins and ends, in the ILR
and CR, respectively. The density falls exponentially along the spiral
arm, where the radial scale length of the galactic disk is used
(depending on morphological type, see Table 1). 

\begin{deluxetable*}{llcccc}
\tablecolumns{6}
\tabletypesize{\small}
\tablewidth{0pt}
\tablecaption{Parameters of the Galactic Models}
\tablehead{\multicolumn{2}{l}{Parameter} &\multicolumn{3}{c}{Value}& {Reference}}
\startdata
  &&Sa&Sb&Sc&\\
\hline
 &&\multicolumn {3}{c}{\it Axisymmetric Components} & \\
\hline
\multicolumn{2}{l}{M$_{B}$ / M$_{D}$} & 0.9& 0.4& 0.2 & 1,2 \\
\multicolumn{2}{l}{M$_{D}$ / M$_{H}$ $^{\rm 1}$}&0.07 &0.09 & 0.1  &  2,3  \\
\multicolumn{2}{l}{Rot. Velocity ($\kms$)}& 320&250 &  170& 4  \\
\multicolumn{2}{l}{ $(V_D/V_{Rot})_{2R_h}$}  &0.51 & 0.65&0.70&\\
\multicolumn{2}{l}{M$_{D}$ ($10^{11}$M$_\odot$)}   &1.28 & 1.21& 0.51& 3  \\
\multicolumn{2}{l}{M$_{B}$ ($10^{11}$M$_\odot$)}  & 1.16& 0.44& 0.10& $M_{B}/M_{D}$ based \\
\multicolumn{2}{l}{M$_{H}$ ($10^{11}$M$_\odot$)}     &16.4 &12.5 & 4.8 & $M_{D}/M_{H}$ based \\
\multicolumn{2}{l}{Disk scale-length ($\kpc$)} & 7&5 & 3 & 1,3\\
 \multicolumn{2}{l}{b$_1$ $^2$ (kpc)}  & 2.5 & 1.7 &1.0& \\
    \multicolumn{2}{l}{a$_2$ $^2$ (kpc)} &7.0&5.0 &5.3178&\\
     \multicolumn{2}{l}{b$_2$ $^2$ (kpc)}&1.5&1.0&0.25&\\
       \multicolumn{2}{l}{a$_3$ $^2$ (kpc)} &18.0&16.0&12.0&\\
\hline
 &&\multicolumn {3}{c}{\it Spiral Arms}&  \\
\hline
\multicolumn{2}{l}{locus}             & \multicolumn{3}{c}{Logarithmic } & 5,9,10\\
\multicolumn{2}{l}{arms number}       & \multicolumn{3}{c}{2} & 6\\
\multicolumn{2}{l}{pitch angle ($\deg$)}       & 8-40& 9-45& 10-60& 4,7 \\
\multicolumn{2}{l}{M$_{sp}$/M$_{D}$}& \multicolumn{3}{c}{1-5\%} & 9  \\
\multicolumn{2}{l}{scale-length ($\kpc$)}    & 7&5& 3  & disk based\\
\multicolumn{2}{l}{$\Omega_{sp}$ ($\kmskpc$) }&-30 &-25 & -20& 5,8 \\
\multicolumn{2}{l}{ILR position ($\kpc$)}    &3.0 &2.29 & 2.03& \\
\multicolumn{2}{l}{CR position ($\kpc$)}  & 10.6& 11.14 & 8.63 & \\
\multicolumn{2}{l}{inner limit   ($\kpc$) }   & 3.0& 2.29& 2.03& $\sim$ILR position based\\

\multicolumn{2}{l}{outer limit ($\kpc$)}  & 10.6& 11.14& 8.63 & $\sim$CR position based\\ 

\label{tab:parameters}
\enddata

\tablenotetext{1} { Up to 100 kpc halo radius.} 
\tablenotetext{2}{b$_1$, a$_2$, b$_2$, and a$_3$ are scale lengths.}

\tablerefs{ (1)~Weinzirl \et 2009; 
                 (2)~Block \et 2002;
                 (3)~Pizagno \et 2005;
                 (4) ~Brosche 1971; Ma \et 2000; Sofue \& Rubin 2001;
                 (5)~Grosb\o l \& Patsis 1998;
                 (6)~Drimmel \et 2000; Grosb\o l \et 2002; Elmegreen \& Elmegreen 2014;
                 (7)~Kennicutt 1981;
                 (8)~Patsis \et 1991; Grosb\o l \& Dottori 2009; 
                      Egusa \et 2009; Fathi \et 2009; Gerhard 2011;
                 (9)~Pichardo \et 2003;
                (10)~Seigar \& James 1998; Seigar \et 2006;
}
\end{deluxetable*}

The mass assigned to build the spiral pattern is subtracted from the disk mass to keep the
given model invariable in mass. PERLAS is a more realistic potential
since it is based on a density distribution and considers the force
exerted by the whole spiral structure, obtaining a more detailed shape
for the gravitational potential, unlike a two-dimensional local arm
such as the tight winding approximation (TWA) represented for a simple
cosine function.

Spiral arms nature is a matter of discussion nowadays, particularly
their long-lasting or transient nature. We have performed experiments
with constant, transient, gradual and sudden presence of the spiral
arms. Although the growth rate is an unknown parameter in galaxies, we
have considered different cases to test. On one hand we produce a set
of experiments where the total mass of the spiral arms is introduced
at once ($t=0$ Gyr). The second set of experiments, inserts the spiral
arms mass linearly in a timelapse of 1 \Gyr. And a third set of
experiments for which the spiral arms are simulated as transient, they
vanish and grow with a given periodicity.
  
\subsection{Initial Conditions and Equilibrium of a Stellar Disk}
\label{sec:IC}

The initial conditions set-up follows the Miyamoto-Nagai density
profile we are imposing. This to avoid transient effects induced by
differences between the initial particle distribution and the imposed
disk potential. In this manner, the initial condition for the stellar
disk is given by

\begin{equation}
\label{eq:MNdensity}
\rho_{{\small MN}}=\\
\frac{b_2^2M_d}{4\pi}\frac{a_2R^2+(a_2+3\sqrt{z^2+b_2^2})(a_2+\sqrt{z^2+b_2^2})^2}{(R^2+(a_2+\sqrt{z^2+b_2^2})^2)^{5/2}(z^2+b_2^2)^{3/2}},
\end{equation}

where $M_d$ is the mass of the galaxy disk, $a_2$ and $b_2$ are the
radial and vertical scale-length, respectively. This three parameters
span a range of values in our simulations in order to capture
different galaxy morphologies and kind of spiral arms.

To distribute the particles according to the Miyamoto-Nagai
  density law we solved equation (\ref{eq:MNdensity}) with a root
  finder method.  This is done by expressing the density $\rho(R,z)$
  in terms of the ratios $\rho(R,0)/\rho(0,0)$ and
  $\rho(R,z)/\rho(R,0)$, which provides us an equation for $R$ and $z$
  in terms of the density. The value of these ratios ranges from $0$
  to $1$, therefore we can explore all the possible values of the
  density with a random function and solve it for R and z.

To assign velocities to the particles we follow the strategy proposed
by \citet{Hernquist1993} where velocities are distributed by an
approximation using moments of the collisionless Boltzmann equation
plus the epicycle approach.

Thus we proceeded as follow: once the density profile has been
established, it is necessary to obtain the rotation velocity. This can
be derived from $\Phi$, the gravitational potential of the model

\begin{equation}
\label{eq:omega}
\Omega_c(R) = \left(\frac{1}{R}\frac{\partial\Phi}{\partial R} \right)^{1/2}
\end{equation}
and $v_c=R\Omega_c(R)$, so the circular velocity is given by

\begin{equation}
\label{eq:vc}
v_c(R) = \left(R\frac{\partial\Phi}{\partial R}\right)^{1/2}.
\end{equation}

Once known $\Omega_c$ at any radii we obtain

\begin{equation}
\label{eq:k}
\kappa = \left( 4\Omega_c^2 + R\frac{d\Omega_c^2}{dR} \right)^{1/2},
\end{equation}

known as the epicyclic frequency, necessary to calculate the velocity
dispersion at $R$ and to correct for the asymmetric drift.

To achieve the requirement for the stellar disk to be in equilibrium,
it is necessary to introduce a given dispersion in the velocity as a
function of $R$. The velocity dispersions in the three polar
coordinates are

\begin{equation}
\label{eq:sigmaR}
\sigma_R = 3.358\frac{\Sigma(R)Q}{\kappa}
\end{equation}

\begin{equation}
\label{eq:sigmaphi}
\sigma_{\phi} = \frac{1}{2}\frac{\sigma_R\kappa}{\Omega_c}
\end{equation}

\begin{equation}
\label{eq:sigmaz}
\sigma_z = \sqrt{\pi G\Sigma b_2},
\end{equation}
where $\kappa$ is the epicyclic frequency, $\Sigma(R)$ the surface
density, $b_2$ is the vertical scale-lenght of the disk, and $Q$ is
the known Toomre parameter. According to \citet{Toomre1964}, local
stability requires $Q > 1$, we choose $Q=1.1$ and found this value to
be sufficient for the three galaxy types. In this way the velocity
dispersion depends on the mass of the components that form each
galaxy.

The asymmetric drift correction is defined as
\begin{equation}
\label{eq:asymmetricdrift}
\left\langle v_{\phi} \right\rangle^2 = v_c^2 - \sigma_{\phi}^2 - \sigma_R^2 \left( -1 - 2\frac{R}{\Sigma} \frac{\partial\Sigma}{\partial R} \right),
\end{equation}

is a correction that has to be implemented in the set up for the
initial conditions given the fact that stellar orbits are not in
general in circular orbits, instead, orbits follow epicycles around a
guiding point at the position of the circular orbit and these
epicycles are characterized by the epicyclic frequency $\kappa$.

Finally the particles are distributed in the velocity space as

\begin{eqnarray}
\label{eq:velocities}
v_{\phi} = \left\langle v_{\phi} \right\rangle \pm x\sigma_{\phi} \nonumber\\
v_R = \left\langle v_R \right\rangle \pm x\sigma_R \\
v_z = \left\langle v_z\right\rangle \pm x\sigma_z \nonumber
\end{eqnarray}
where $x$ is a random number between 0 and 1, $\left\langle v_{\phi}
\right\rangle$ is given by Eq. (\ref{eq:asymmetricdrift}) and the
average radial and vertical velocities are taken as $\left\langle v_R
\right\rangle = \left\langle v_z \right\rangle=0$.

\subsection{Dispersion analysis}
The disk heating is often referred as the increase in the velocity
dispersion over the lifetime of a star. Any disk thickening is then
related to an increase in the vertical velocity dispersion of the disk
stars. In this paper we analyze the spiral arms effects on the stellar
disk, based on the study of the vertical velocity dispersion and its
dependence with the parameters that characterize the spiral pattern.

The vertical velocity dispersion $\sigma_z$, is then calculated in the
simulations by dividing the plane $z=0$ into $1\kpc$ bins and
computing, as usual, the squared root of the averaged squared vertical velocity for all the
particles that fall into a given bin. This provide us the vertical
velocity dispersion as a function of $R$. In order to establish the
contribution of the spiral arms to the disk thickness we also compute
$\sigma_z$ as a function of time by measuring the velocity dispersion
at a fixed radius $R$ for every time code unit (in this case, every
$100\Myr$ across $5\Gyr$ in the simulation).

\subsection{Control Simulations: Testing the Initial Conditions Equilibrium}
As described in section \ref{sec:IC}, the first goal is to build an
initial stellar disk in equilibrium to be sure that any change seen in
radial and vertical velocity dispersion is strictly due to the
interaction of the spiral arms with the stellar disk and not
originated by a spurious non-relaxed initial condition set-up.

\begin{figure}
\begin{center}
\includegraphics[width=9cm]{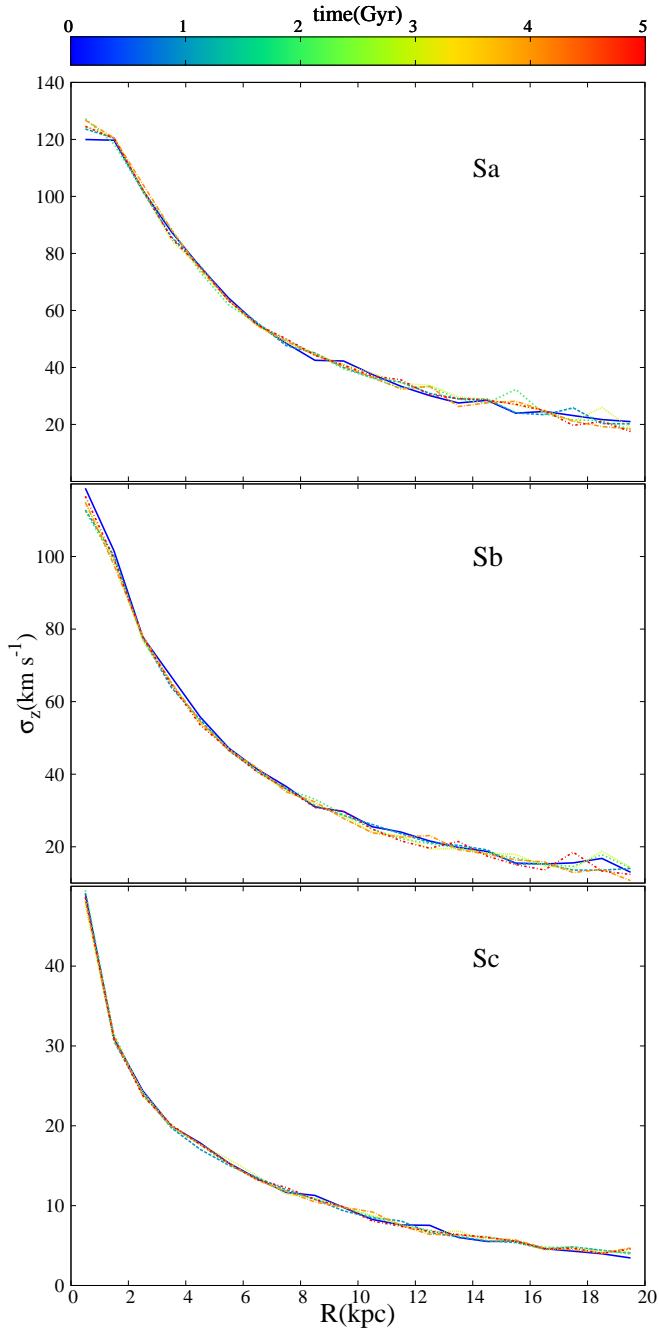}\\
\end{center}
\caption{Test for the initial conditions equilibrium: velocity
  dispersion of the stellar disk in the axisymmetric potential only,
  as a function of $R$, along $5\Gyr$ time evolution for each galaxy
  type.}
\label{fig:axisymmetric}
\end{figure}

For this test a control simulation was produced with only the
axisymmetric components for the potential model. Stars are run by
$5\Gyr$, for all galaxy types. Figure \ref{fig:axisymmetric} shows
$\sigma_z$ as a function of $R$ at different stages in the temporal
evolution for an Sa, Sb, and Sc galaxy.

From the figure it is clear that the vertical velocity dispersion do
not evolve or deviate from the initial dispersion, as expected for a
disk in equilibrium with the axisymmetric potential.

\section{Results} \label{results}
We present in this section a set of controlled experiments to study
the dynamical heating of disks on spiral galaxies considering the
spiral arms as the driver. The general purpose is to shed some light
on the relative importance of these large-scale structures to other
sources of dynamical heating in different morphological types. The
experiments include studies of different structural and dynamical
parameters of the spiral arms such as, different pitch angles, total
spiral arms masses, angular speeds, transient, and one final case
modeling the Milky Way Galaxy (with preliminary results that will be
better developed in a future work).

\subsection{Dependence of the disk heating induced by spiral arms with galaxy morphology}
Spiral galaxies present a wide variety on morphological types, from
massive bulge-dominated galaxies to practically bulgeless disks,
spanning a wide range of values for the parameters that characterize
different galaxy types.

\begin{figure}[H]
\begin{center}
\includegraphics[width=9cm]{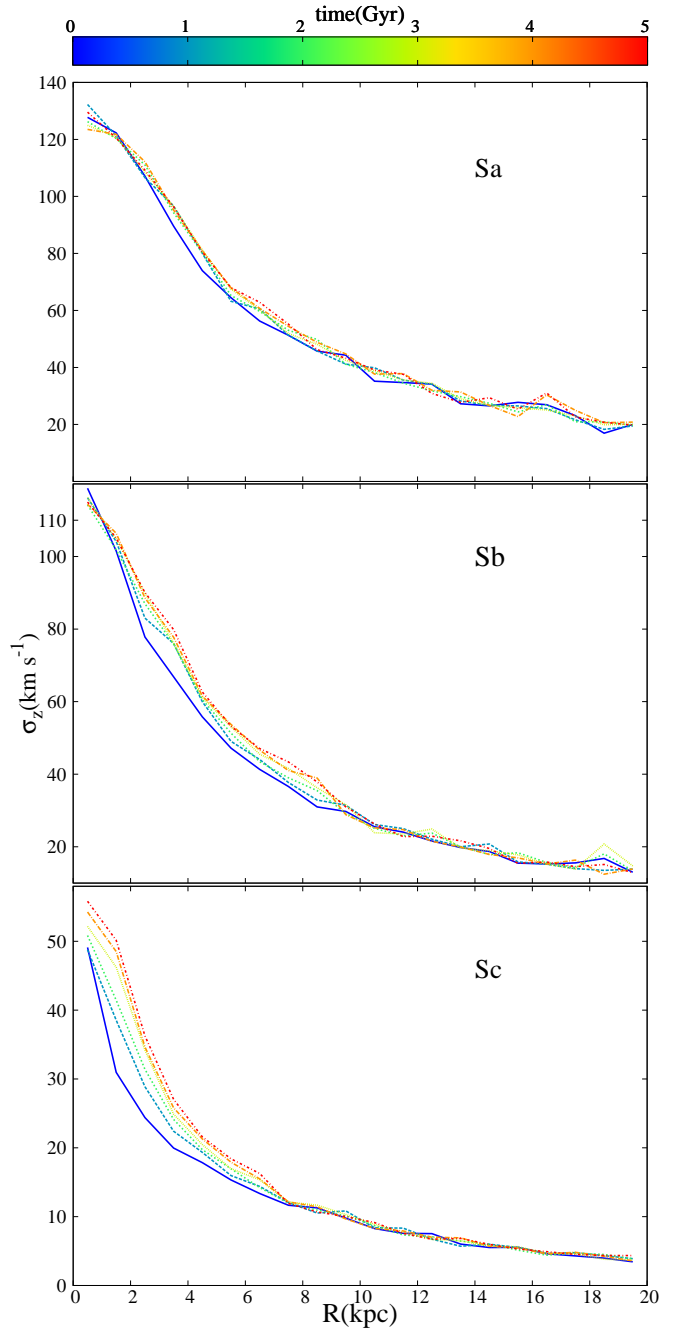}
\end{center}
\caption{As in Figure \ref{fig:axisymmetric}, but including now the
  spiral arms contribution to the potential. The vertical velocity
  dispersion is plotted as a function of $R$ for $5\Gyr$ time
  evolution.}
\label{fig:morphology}
\end{figure}

Figure \ref{fig:morphology} shows the velocity dispersion, $\sigma_z$,
as a function of the galactocentric radius, $R$, at different times in
the simulation for our galactic models: Sa, Sb and Sc (introduced in
Section \ref{model}). For these three simulations the mass of the
spiral arms, $M_{arms}$, is $5\%$ of the total disk mass with a Pitch
angle of $40^{\circ}$ for the Sa galaxy, $45^{\circ}$ for the Sb
galaxy, and $40^{\circ}$ for the Sc galaxy. For these experiments we
have employed the largest spiral arms masses and pitch angles, for
plausible (non-fully chaotic) galactic models, to identify clearly
spiral arms effects, if any. The three plots show a distinct increase
in the vertical velocity dispersion caused by the spiral arms. Also
from figure \ref{fig:morphology} it is clear that the change in
$\sigma_z$ with respect to the initial dispersion is smaller for the
Sa galaxy and grows with the morphological type, being much larger for
the Sc galaxy model.

\begin{figure*}
\begin{center}
\includegraphics[width=18cm]{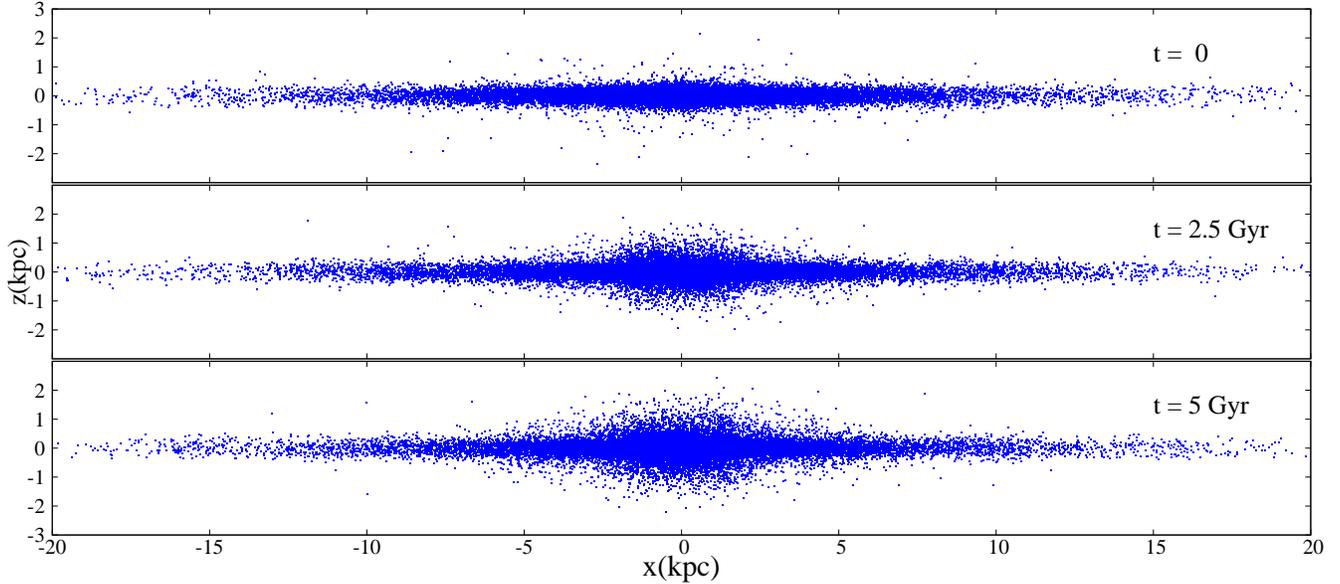}
\end{center}
\caption{Edge on disks show the evolution of stellar orbits at three
  different times (0, 2.5 and 5 Gyr). The model corresponds to an Sc
  galaxy. The thickening of the disk due to the spiral arms presence
  is clear.}
\label{fig:edgeon}
\end{figure*}

The dependence of the effect of the spiral arms with the morphology is
such that for an Sc galaxy the effect is evident in the spatial
distribution of the stellar disk particles. Figure \ref{fig:edgeon}
shows the $x-z$ projection of the stellar distribution plotted at
$t=0$, $t=2.5\Gyr$, and $t=5\Gyr$. 

Also, a thickening of the disk is
discernible during the orbital evolution when comparing with the
initial distribution.

\begin{figure}
\begin{center}
\includegraphics[width=9cm]{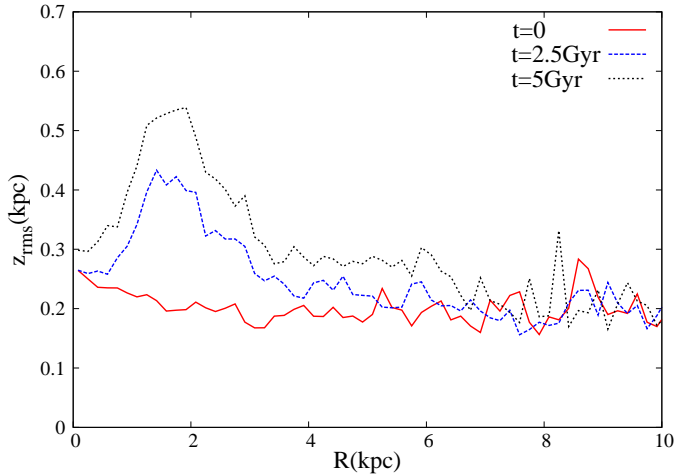}\\
\end{center}
\caption{Disk thickness $z_{rms}$ as a function of $R$ for $t = 0$, $2.5\Gyr$ and $5\Gyr$.}
\label{fig:thickness}
\end{figure}

The thickness can be quantified by computing the root mean square of
the coordinate $z$, i.e. $z_{rms} = \sqrt{<z^2>}$. Figure
\ref{fig:thickness} shows the thickness as a function of $R$ for the
stellar disks in Figure \ref{fig:edgeon}.

From these first set of experiments, we conclude that the sharpest
effect on the velocity dispersion, is present on the latest
morphological types. It is worth mentioning here that, although
  we are separating the models in Sa, Sb and Sc galaxies, with strong
  gaps in between the different models in the initial scale height,
  the results on the disk heating driven by spiral arms presented
  here, are more general, i.e. the observed heating results
  significant in thinner disks, which, as a consequence, has
  implications on galaxy types, in this case, particularly on later
  types.

For the earliest type (Sa), the isolated effect of spiral arms
corresponds to a maximum increment of 7\% percent of the initial
velocity dispersion, this considering the most massive and the largest
pitch angles possible to produce plausible galactic models. Likewise,
for the intermediate galactic type (Sb), the isolated effect of spiral
arms corresponds to a maximum of increment 20\% percent of the initial
velocity dispersion, again considering the most massive and the
largest pitch angles possible to produce plausible galactic
models. Finally, for the latest type (Sc), the isolated effect of
spiral arms corresponds to a maximum of 62\% percent of the initial
velocity dispersion for this example, where we have not used the
maximum plausible parameters for the spiral arms. When the maximum
pitch angle and mass is employed, the percentage goes up to almost
$90\%$ of the initial velocity dispersion.

Since a visible effect in the vertical dispersion when spiral arms are
included is considerably larger for the latest galactic types,
compared with early and intermediate types, in the next sections we
concentrate on a more detailed study of the disk thickening focusing
only in the late type galaxies.

\subsubsection{The pitch angle effect}
\label{PitchAngle}

The pitch angle is one of the most influential structural parameters
that characterize spiral patterns. In this section a range of values
is explored in order to quantify the dependence of the increment in
the vertical velocity dispersion with pitch angle in the most affected
galactic models, that are the latest types.

\begin{figure} 
\begin{center}
\includegraphics[width=9cm]{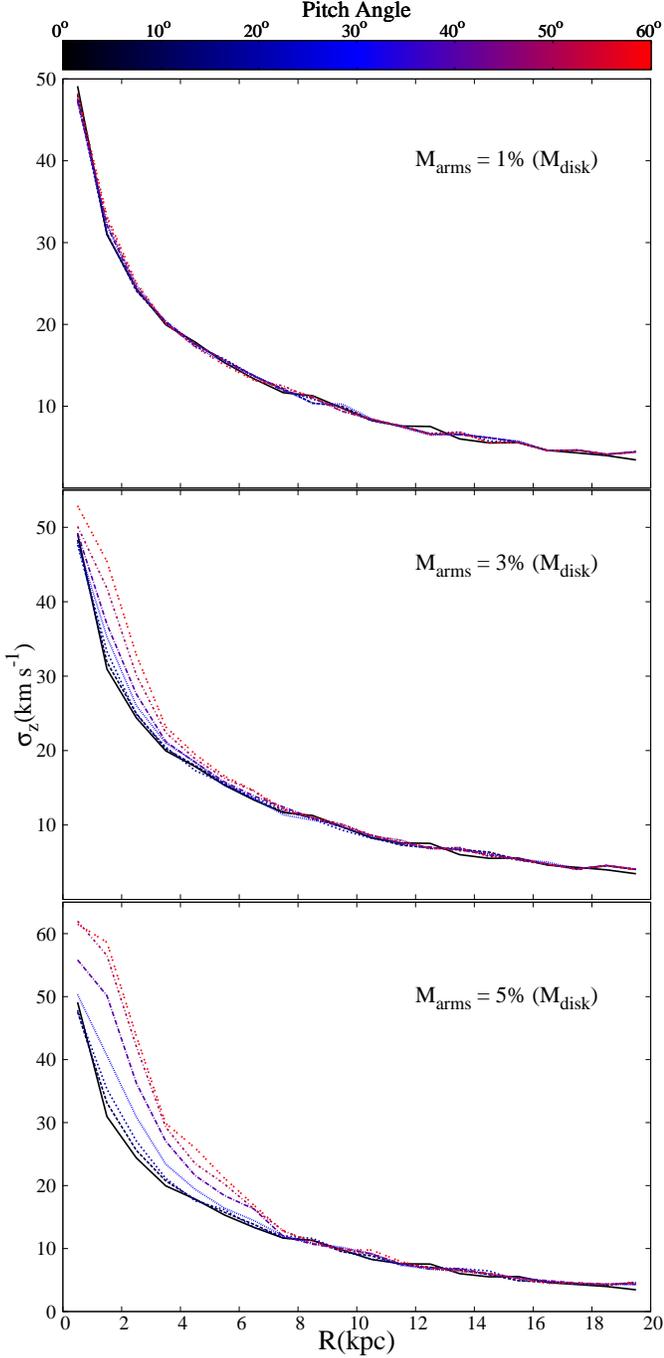}
\end{center}
\caption{Final velocity dispersion after a $5\Gyr$ evolution as a
  function of radius for spiral arm masses of $1\%$, $3\%$, and $5\%$
  of the total disk mass.}
\label{fig:pitchangle}
\end{figure}

Figure \ref{fig:pitchangle} shows three plots of $\sigma_z$ vs $R$,
where the mass of the spiral arms is set to a constant for each plot
and the pitch angle vary according to Table 1. Each plot has the
initial dispersion curve $\sigma_z(R,t=0)$ and the dispersion after a
$5\Gyr$ evolution $\sigma_z(R,t=5\Gyr)$ for each pitch angle value.

From Figure \ref{fig:pitchangle}, it is clear that regardless the mass
of the spiral arms, the vertical velocity dispersion increases notably
with the pitch angle. The less massive the spiral arms the smaller is
their effect in general, as expected. Indeed, for spiral arms masses
smaller that $\sim 1\%$ of the disk, the contribution of spiral arms
to the dynamical heating becomes negligible.

\subsubsection{The spiral arm mass effect}

\begin{figure}[H] 
\begin{center}
\includegraphics[width=9cm]{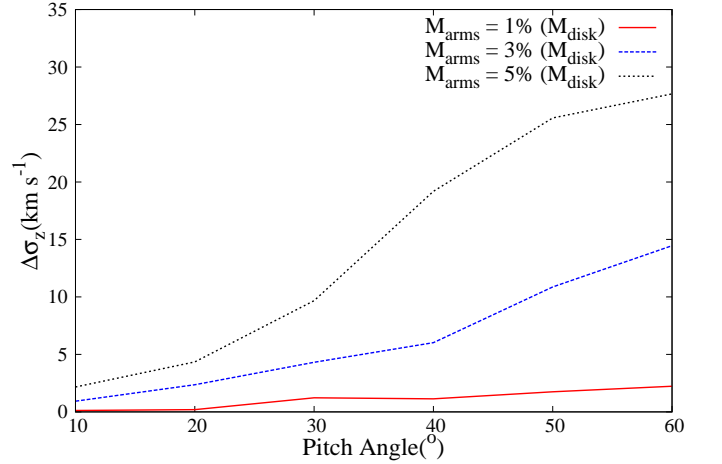}\\
\end{center}
\caption{Difference in the vertical velocity dispersion
  $\Delta\sigma_z$ (at t=0 and t=5Gyr) vs the pitch angle, plotted for
  the three masses of the spiral arms, as a scaling factor.}
\label{fig:mass}
\end{figure}

As it is shown in the previous section, the effect of the pitch angle
can be significant to the disk thickening. It also results intuitively
clear that this also scales with the mass of the spiral arms. To
address this point, we produced several experiments condensed in
Figure \ref{fig:mass}, that show the disk thickening dependence on the
spiral arms mass and the pitch angle together. For this purpose, first
we identified the radius $R_{max}$ at which occurs the maximum
difference between the final and initial vertical velocity
dispersion. We find that this occurs at approximately
$R_{max}=1.5\kpc$ for this model. We then keep $R$ constant at that
value and plot $\Delta\sigma_z$ there as a function of the pitch
angle. This will show us the tendency seen in the previous
section. Now repeating this for the three spiral arms masses employed
will show how $\Delta\sigma_z$ scales with this second parameter.

Figure \ref{fig:mass} shows $\Delta\sigma_z$ as a function of the
pitch angle for three different masses of the spiral arms. Here it is
clear that both parameters, the pitch angle and the mass of the spiral
arms, affect considerably the disk thickening effect driven by the
spiral structure.

We produce with these results an empirical functional relation between
$\Delta\sigma_z$ and the Pitch Angle. The fit of data plotted in
figure \ref{fig:mass} is made by noting that $\Delta\sigma_z$,
increases slowly at small angles, then the slope of the curve grows
with the angle and flattens for the largest values of the pitch
angle. This behavior could be interpreted as a saturation effect of
$\Delta\sigma_z$ after a certain time, this time being shorter for
higher angles and masses. The saturation effect is also seen in a
$\sigma$ - $time$ relation
\citep{Seabroke2007,Soubiran2008,Calberg1985}, where the dispersion
remains constant after $\sim$ $5\Gyr$. Based on this observation, a
nice fit to the results would be a Boltzmann sigmoidal function which
is characterized by displaying a progression from small beginnings
that accelerates and approaches a climax over the independent
variable. The Boltzmann sigmoidal function, for this particular case,
is defined by

\begin{equation}
\label{eq:sigmoidal}
\Delta\sigma_z = A_2 + \frac{(A_1-A_2)}{1+exp((x-x_0)/d)}
\end{equation}

\noindent where:

 \noindent $x$   = Pitch Angle ($^{o}$) \\
 $x_0$ = center ($^{o}$)\\
 $d$   = width ($^{o}$)\\
 $A_1$ = initial $\Delta\sigma_z$ value ($\kms$)\\
 $A_2$ = final $\Delta\sigma_z$ value ($\kms$)\\

The center $x_0$ is the pitch angle at which $\Delta\sigma_z$ is halfway between $A_1$ and $A_2$. The width $d$ is related with the steepness of the curve, with a larger value denoting a shallow curve.

\begin{figure}[H]
\begin{center}
\includegraphics[width=9cm]{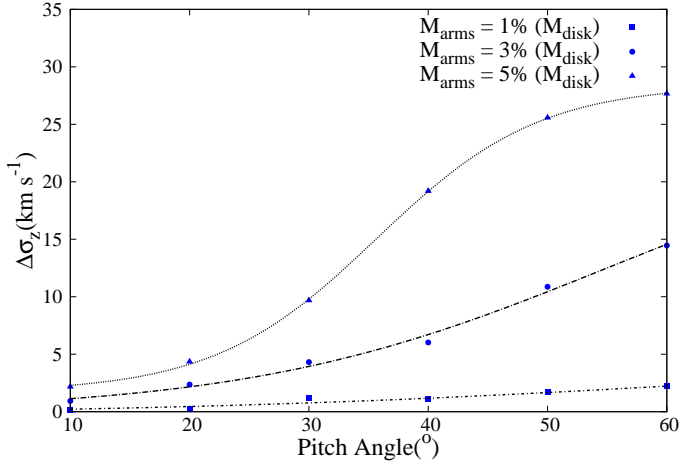}
\end{center}
\caption{Fit of the $\Delta\sigma_z$ - pitch angle relation with
  the Boltzmann sigmoidal function for the three masses of the spiral
  arms.}
\label{fig:fitmass}
\end{figure}

Figure \ref{fig:fitmass} shows the fit of the sigmoidal function
(Eq. \ref{eq:sigmoidal}) to the data. We see that this function
reproduces well the behavior of $\Delta\sigma_z$ with the pitch angle
for the three masses of spiral arms used in our simulations.

Taking Eq. (\ref{eq:sigmoidal}) as the functional form for the
$\Delta\sigma_z$ - pitch angle relation, in Table \ref{tab:fit}, we
summarize the parameters that describe the fits plotted in figure
\ref{fig:fitmass} for the three spiral arms masses.

\begin{deluxetable*}{ccccc}
\tabletypesize{\footnotesize}
\tablecolumns{5}
\tablewidth{0pt}
\tablecaption{ $\Delta\sigma_z$ - Pitch Angle Fit Details 
\label{tab:fit}}
\tablehead{$M_{arms}$ ($\%M_{disk}$) & $x_0$ $(^{o})$ & $d$ $(^{o})$ & $A_1 $ $(\kms)$  & $A_2$ ($\kms$) }
\startdata
$1\%$      &  64.73  &  24.05  &  -0.33    & 5.32   \\
$3\%$      &  54.73  &  14.95  &  -0.057   & 24.84  \\
$5\%$      &  35.68  &  6.82   &   1.69    & 28.44  \\
\enddata
\tablecomments{Parameters that combined with equation
  \ref{eq:sigmoidal}, describe the $\Delta\sigma_z$ - pitch angle
  relation. Center $x_0$, width $d$, initial $\Delta\sigma_z$ value
  $A_1$, and final $\Delta\sigma_z$ value $A_2$. For the three spiral
  arms masses (in percentage of the disk mass $M_{disk}$).}
\end{deluxetable*}

\bigskip
\bigskip
\centerline{\it Gradually Increasing the Mass of the Spiral Arms}

As explained before, the spiral structure in our galaxy potential
models is imposed and fully introduced since the starting point of the
simulation. However, one might wonder if spurious effects (such as a
drastic dispersion increase in the early stages of the simulation) are
introduced with this method; also, in real galaxies the birth and
death of spiral arms is most probable not a sudden process.

\begin{figure}
\begin{center}
\includegraphics[width=9cm]{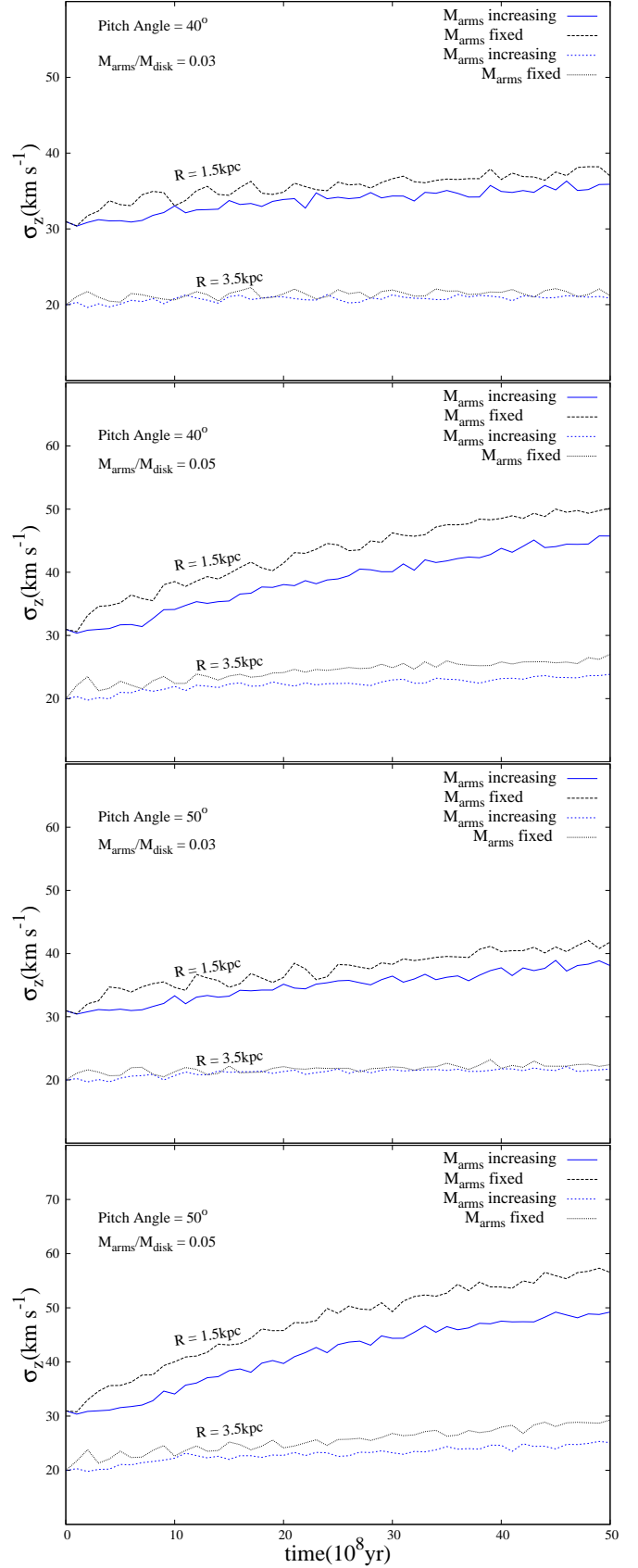}\\
\end{center}
\caption{$\sigma$ - $t$ relation that shows, regardless the radial
  position, a continuous heating for both kind of imposed spiral arms:
  those that grow with time (blue lines) and those totally formed
  since the beginning (black lines).}
\label{fig:gradually}
\end{figure}

In order to explore these two scenarios, we prepared a set of
representative simulations modifying the model in such a way that the
arms are allowed to have a growing period until the total assigned
mass is reached. For this set of simulations we started with an Sc
galaxy with different combinations of values for the spiral arms mass
and the pitch angle. We chose two radial positions and measured in
there the evolution of $\sigma_z$ with time as showed in Figure
\ref{fig:gradually}. In this figure the temporal evolution of
$\sigma_z$ for the sudden full mass arms is compared with a model with
linear ($1\Gyr$ period) growing mass. It is clear that there is a
shift in the dispersion achieved at the end of the simulation, but
most important, for both kinds of imposed spiral arms the velocity
dispersion increases at roughly the same rate.

With this simple experiment we are not pretending to capture the great
variety of processes that lead to the formation of a real spiral
pattern, nor their timelapse of full action, but this is still likely
a better approximation than a sudden action of spiral arms. With this
excercise we notice however that, for any radial position, the
vertical dispersion is only shifted when using spiral arms that have a
growing period. This do not suppose important differences with our
previous simulations or results derived from them.

\subsubsection{Varying the Angular Velocity of the Spiral Pattern}
The vertical heating of the stellar disk produced by the spiral arms
depends on the parameters of the nonaxisymmetric structure,
particularly for the late galaxy types. By varying the pitch angle and
the mass of the arms we have found a correlation between this
parameters and the thickness of the stellar disk.

\begin{figure*}
\begin{center}
\includegraphics[width=19cm]{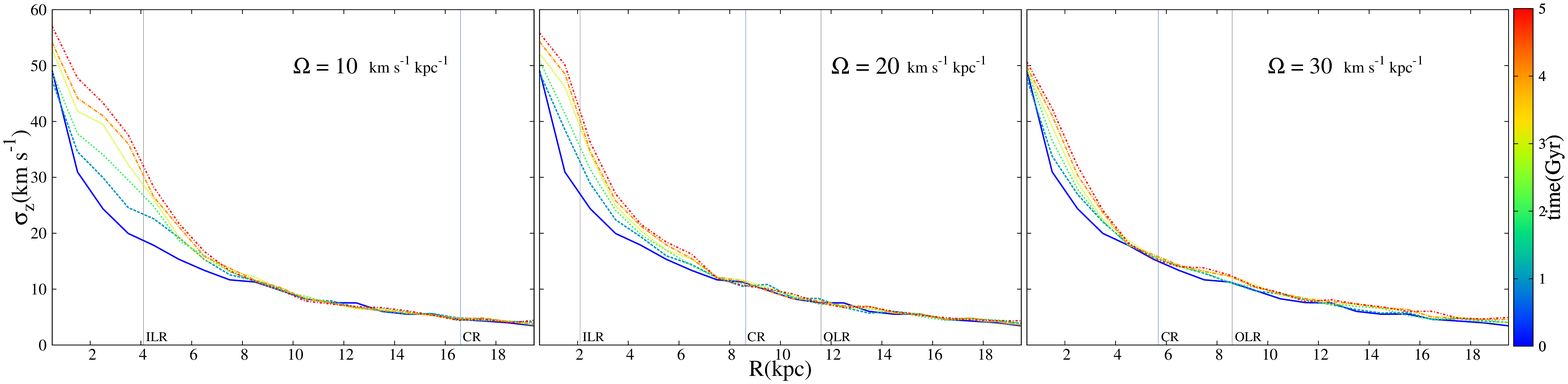}\\
\end{center}
\caption{Time evolution of the vertical velocity dispersion for three
  different spiral pattern angular speeds in an Sc galaxy. This
  comparison shows that slower rotating spiral arms heat more
  efficiently the stellar disk.}
\label{fig:velang}
\end{figure*}

In this section, we study now the pattern speed effect on the disk
vertical heating. For this purpose, we ran a set of simulations with
different values of the spiral pattern for the latest types of
galaxies: Sb and Sc, that are more clearly affected for the structural
parameters of the arms. In Figure \ref{fig:velang} we show first the
evolution of the vertical velocity dispersion $\sigma_z$ over time up
to $5\Gyr$ for different values of the pattern speed, $\Omega$, for an
Sc galaxy. There seems to be a clear relation between the angular
velocity and the dynamical heating. The velocity dispersion increases
for slower rotating patterns. 

\begin{figure}[H]
\begin{center}
\includegraphics[width=9cm]{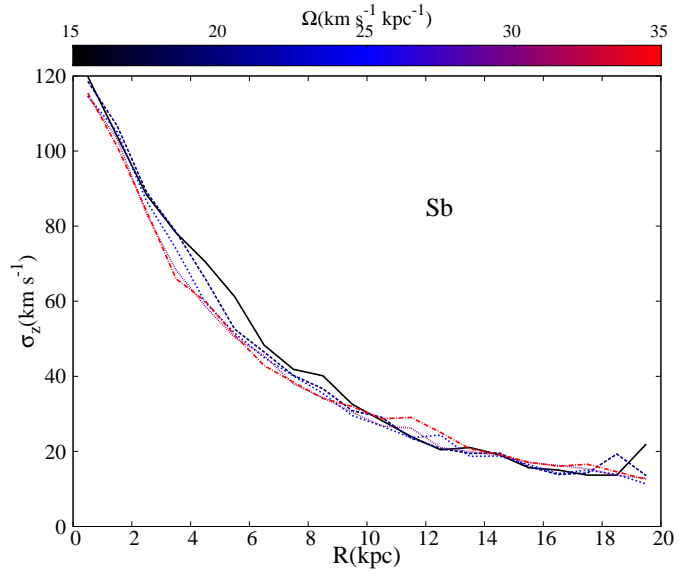}\\
\end{center}
\caption{Final velocity dispersion after a $5\Gyr$ evolution for
  different spiral pattern angular speeds in an Sb galaxy.}
\label{fig:CompareVelAngSb}
\end{figure}

\begin{figure}[H]
\begin{center}
\includegraphics[width=9cm]{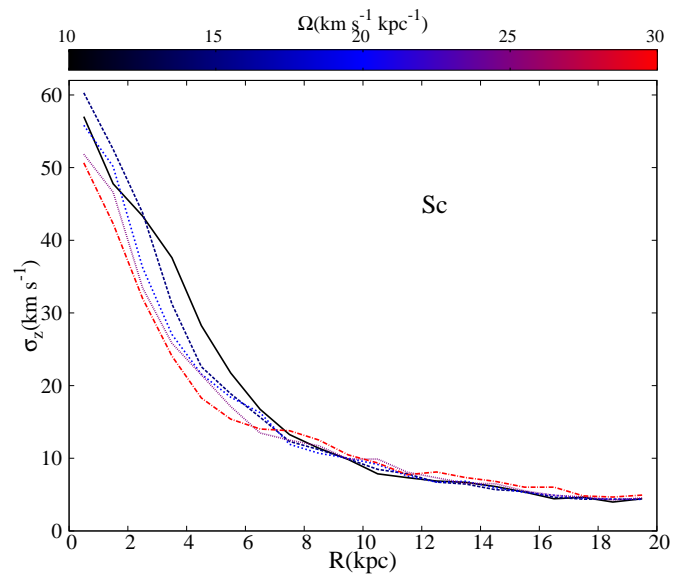}\\
\end{center}
\caption{Final velocity dispersion after a $5\Gyr$ evolution for
  different spiral pattern angular speeds in an Sc galaxy. }
\label{fig:CompareVelAng}
\end{figure}

Figures \ref{fig:CompareVelAngSb} and \ref{fig:CompareVelAng}
summarize the results of the set of simulations when varying the
pattern speed for Sb and Sc galaxies. The final stage of $\sigma_z$ is
shown across the entire disk for each one of the pattern speeds
used. As expected, the final vertical velocity dispersion and disk
thickness are larger for slower rotating arms, independently of the
galaxy type. Smaller values of $\Omega$ allow the spiral arms to heat
more efficiently the stellar disk, likely due to the minor relative
angular velocity between the arms and the stars, this allows the stars
to interact with the potential of the arms for longer periods of time.

Other studies link the heating of the stellar disk and the role of the
pattern speed only to the resonant regions of the spirals \citep{Lynden1972, DeSimone2004}, 
in contrast, in this study we find that the
heating, radial and vertical, occurs along the entire length of the
spiral arms.

On the other hand N-body simulations had shown to develop transient
spiral structure that spans a range of pattern speeds \citep{Sellwood2002, Roskar2012}, 
but as the arms are transient and is not possible to isolate its effect in N-body simulations, 
it results difficult to establish a dependence of the heating on the pattern speed.

The experiments and the model shown here allows us to establish a
relation between the vertical heating of the stellar disk and the
pattern speed.

\subsubsection{$\sigma_z$ - Time Relation}

It is already known that the age and velocity dispersion of stars are
correlated. This has been established from observations in the solar
neighbourhood as well as from numerical simulations
\citep{Holmberg2009, Roskar2013, Martig2014}. The $\sigma$ - $t$
relation shows a smooth, general increase of the velocity dispersion
with time and is best parametrized by a power law with exponents
ranging between $0.2$ - $0.5$ \citep{Gerssen2012}.

We explore the $\sigma$ - $t$ relation in our simulations to find out
if the velocity dispersion in the stellar disk due to the spiral arms
fits with a power law $t^{\alpha}$ and more important, to establish a
range of values for $\alpha$.

To measure the time evolution of $\sigma_z$ in our simulations first
we locate the radius $R_{max}$ at which occurs the maximum increase in
the vertical velocity dispersion. This radius is $R_{max}=1.5\kpc$ and
is the same for all the simulations with the Sc galaxy, independently
of the pitch angle or the mass of the spiral arms. 

\begin{figure}[H]
\begin{center}
\includegraphics[width=9cm]{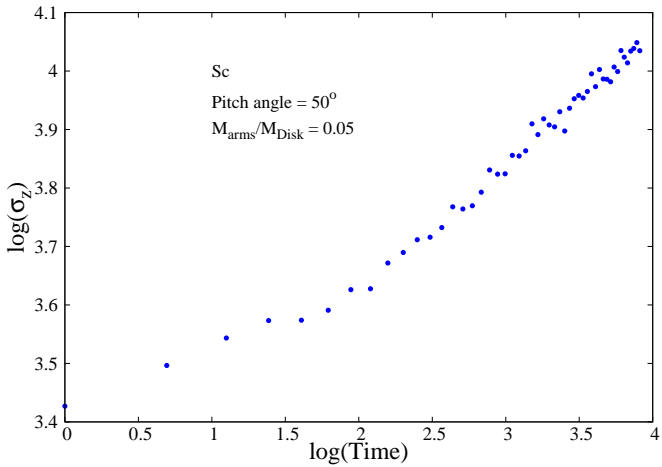}
\end{center}
\caption{log-log plot of the time evolution of $\sigma_z$.}
\label{fig:loglog}
\end{figure}

Since a log-log plot is useful to recognize a possible power law
relationship, Figure \ref{fig:loglog} shows the $log(t)$ -
$log(\sigma_z)$ dependency. The most of the points fall on a straight
line, this reveals a behavior of the form $\sigma_z$ $\propto$
$t^{\alpha}$.

\begin{figure}
\begin{center}
\includegraphics[width=9cm]{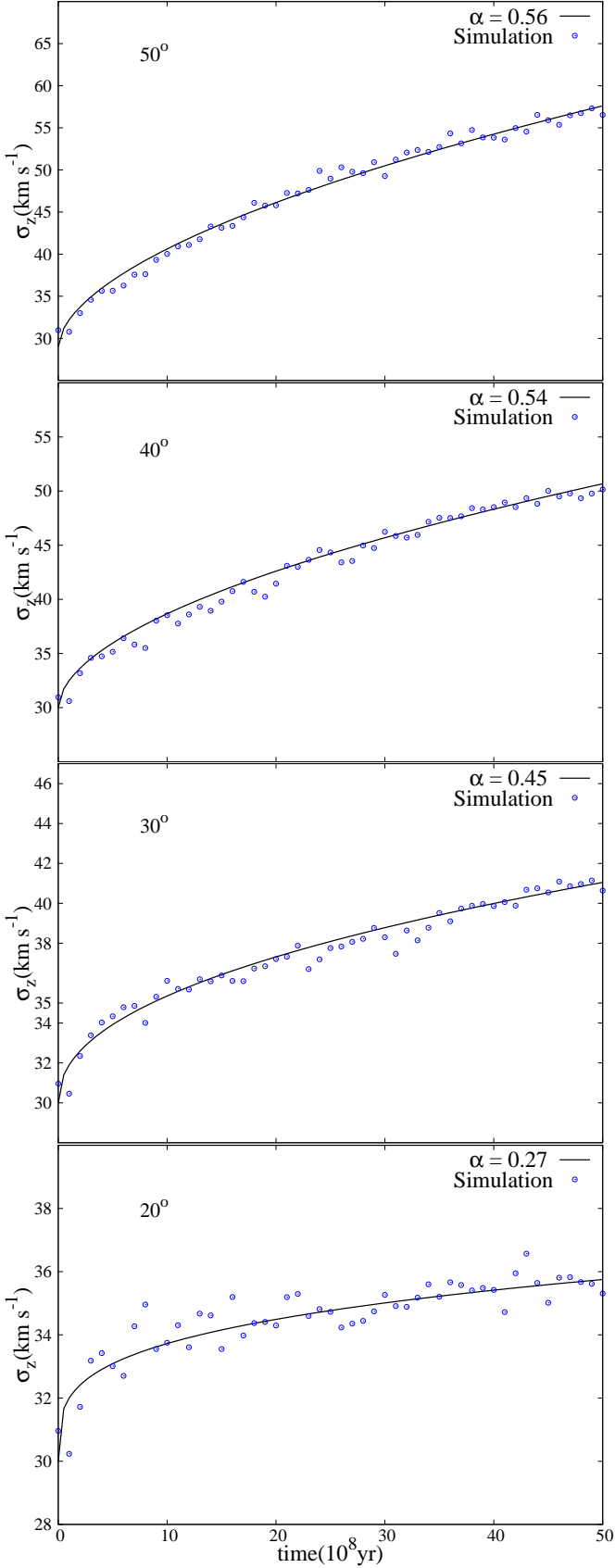}
\end{center}
\caption{Temporal evolution of $\sigma_z$ for a Sc galaxy. The panels
  correspond to four different pitch angles, from top to bottom:
  $50^o$, $40^o$, $30^o$, and $20^o$. The black lines show the best
  fit to the data for a relation of the form $\sigma_z$ $\propto$
  $t^{\alpha}$. }
\label{fig:Scsigmavstime}
\end{figure}

Figure \ref{fig:Scsigmavstime} shows the evolution of $\sigma_z$ with time at
$R=R_{max}$, $M_{arms}/M_{disk}=0.05$ and at different pitch angles:
$20^o$, $30^o$, $40^o$ and $50^o$. The black line in each plot is the
best fit of the data with a power law of the form $\sigma_z$ $\propto$
$t^{\alpha}$.  We made the same analysis for a spiral arm mass of
$M_{arms}/M_{disk}=0.03$, an interesting outcome is that the value of
$\alpha$ is independent of the spiral arm mass. Different masses for
the spiral arms will just change the proportionality constant in the
relation $\sigma_z$ $\propto$ $t^{\alpha}$.  Consequently, $\alpha$
depends only on the pitch angle, and for the angles used in our
simulation $\alpha$ varies within the range $0.27 - 0.56$ for Sc
galaxies.

\begin{figure}[H]
\begin{center}
\includegraphics[width=9cm]{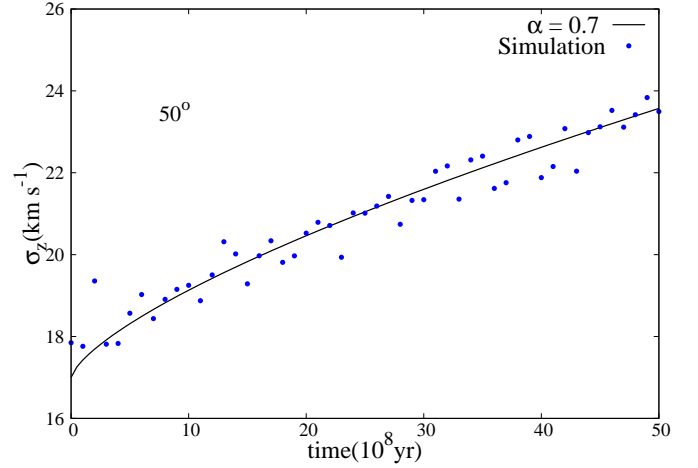}
\end{center}
\caption{Time evolution of $\sigma_z$ measured at half arm length for a Sc galaxy, $50^o$ pitch angles. 
The black line shows the best fit to the data for a relation of the form $\sigma_z$ $\propto$
  $t^{\alpha}$.}
\label{fig:ScsigmavstimeMedioBrazo}
\end{figure}

Although we have presented the time evolution of $\sigma_z$ at $R=R_{max}$, 
it is possible to made the measure at any value of $R$. Figure \ref{fig:ScsigmavstimeMedioBrazo} shows the 
same analysis at a different radial position, and this time corresponds to 
the half of the arm length, $R=4.2 \kpc$. We notice that measuring the time evolution 
of $\sigma_z$ at different radius give us the same power law behavior.

\begin{figure}[H]
\begin{center}
\includegraphics[width=9cm]{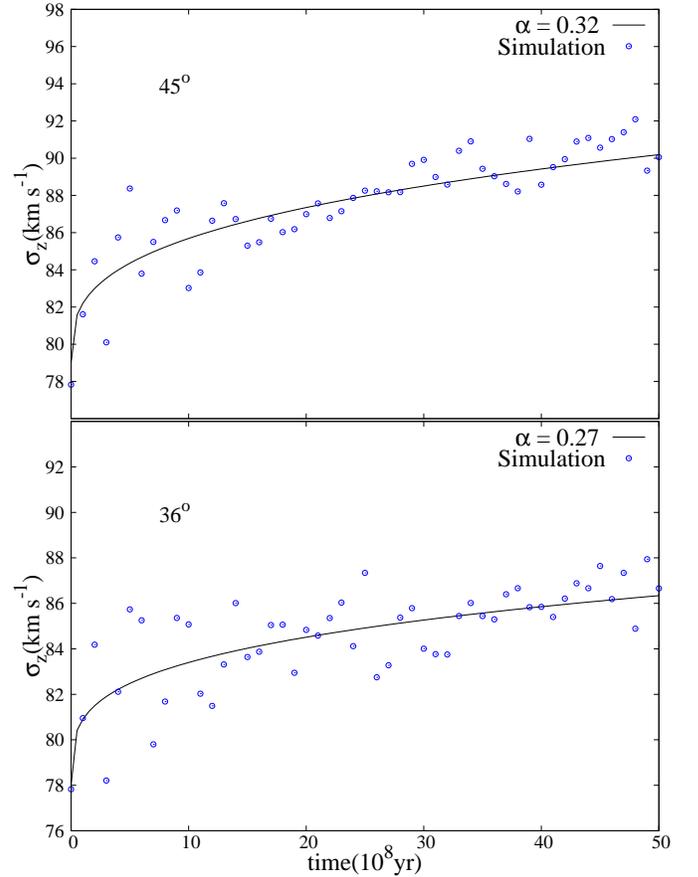}
\end{center}
\caption{Temporal evolution of $\sigma_z$ for an Sb galaxy. The panels
  correspond with two different pitch angles, $45^o$ (top) and $36^o$
  (bottom). The black lines show the best fit to the data for a
  relation of the form $\sigma_z$ $\propto$ $t^{\alpha}$. }
\label{fig:Sbsigmavstime}
\end{figure}

With a similar analysis for Sb galaxies, where $R_{max}=2.5\kpc$, we
find again that a power law provides a nice fit to the data. Figure
\ref{fig:Sbsigmavstime} shows the $\sigma_z$ temporal evolution for
pitch angles of $36^o$ and $45^o$, and the best fits are reached with
values for $\alpha$ of $0.32$ and $0.37$, respectively. Smaller angles
than those give us plots with more scattered points where a power law
fit is not straightforward to obtain, i.e., we are not able to
establish the value of $\alpha$ for angles smaller than $36^o$. For Sb
galaxies we can only provide an upper bound for $\alpha$ of $\sim
0.37$.

Considering that the pitch angles we are employing here represent the
maximum plausible values of pitch angle and spiral arms mass for each
type of galaxy, before chaos dominates, the values of $\alpha$
presented here would represent an upper bound for the contribution of
the spiral arms to the vertical velocity dispersion of stars in each
galaxy type.

\subsubsection{The velocity ellipsoid}
For the the dynamical evolution analysis of the stellar disk we
applied a classic method derived from the distribution of the velocity
dispersions. The axes of such distribution define the known as the
``velocity ellipsoid'', and this is characterized by the two axes
ratios: $\sigma_z/\sigma_R$ and $\sigma_{\phi}/\sigma_R$.

\begin{figure}[H]
\begin{center}
\includegraphics[width=9cm]{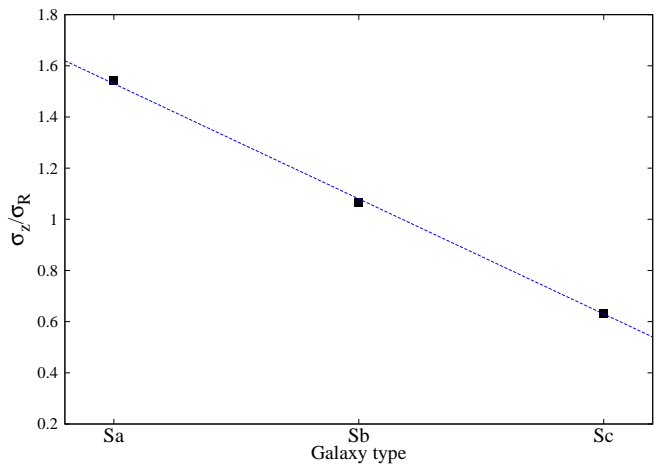}
\end{center}
\caption{Velocity ellipsoid ratio $\sigma_z/\sigma_R$ as a function of
  morphological galaxy type. The dashed line represents a linear
  regresion of the data.}
\label{fig:SzSr}
\end{figure}

The shape of the velocity ellipsoid seems to show a trend in
$\sigma_z/\sigma_R$ with the morphological type
\citep{Gerssen2012}. Given the three types of galaxies we are using in
this work and the parameters given in Table \ref{tab:parameters} that
defines them, we are able to measure the shape of the velocity
ellipsoid and compare it with the morphological type. In Figure
\ref{fig:SzSr} we plot this ratio as measured at the final stage of
our simulations. The value $\sigma_z/\sigma_R$ decrease with galaxy
type and the trend keeps with time.

\begin{figure}[H]
\begin{center}
\includegraphics[width=9cm]{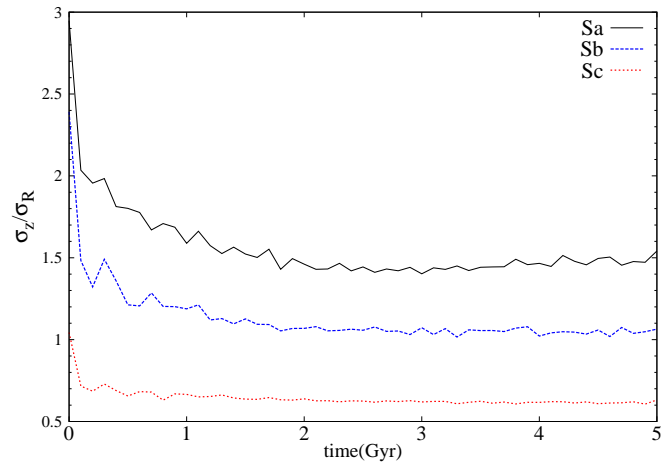}
\end{center}
\caption{Velocity ellipsoid shape at $5\Gyr$ time evolution in our
  simulations for the three galaxy types simulated.}
\label{fig:SzSrtime}
\end{figure}

In Figure \ref{fig:SzSrtime} we follow the temporal evolution for the
shape of the velocity ellipsoid. We see that for Sa and Sb galaxies
$\sigma_z$ is always greater than $\sigma_R$ and the presence of
spiral arms do not seem to alter this tendency. The initial fall in
the curves indicates that although $\sigma_z$ and $\sigma_R$ both
increase with time, $\sigma_R$ does it in a greater proportion than
$\sigma_z$, especially for late type galaxies. This heating in the 
$R$-direction is similar to that seen in Figure \ref{fig:morphology}.

The shape of the velocity ellipsoid evolves towards lower values and
settles down to be nearly constant and fluctuates around some
equilibrium value, a behavior already noticed in other numerical work
\citep{Sellwood2008}.

Looking at the values of the plots in Figure \ref{fig:SzSrtime}, we
see that the velocity dispersions ratio falls to half its initial
value for Sa and Sb galaxies, with $\sigma_z$ always greater than
$\sigma_R$, but the final value of $\sigma_z/\sigma_R$ falls lower
than one for Sc galaxies. This indicates that the relative increase of
$\sigma_R$ for Sc galaxies is greater than that seen on Sa and Sb
galaxies. This means that the heating effect of the spiral arms in Sc
galaxies, apart from being more notorious in the $z$-direction, is
also greater in the $R$-direction compared to the other galaxy types.

\begin{figure}[H]
\begin{center}
\includegraphics[width=9cm]{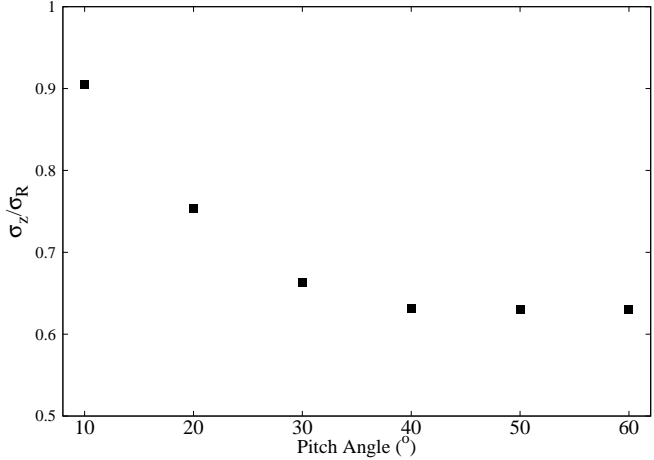}
\end{center}
\caption{Velocity ellipsoid shape, measured after a $5\Gyr$ time
  evolution as a function of the pitch angle in an Sc galaxy.}
\label{fig:ScSzSr50}
\end{figure}

In previous sections we have shown that the vertical heating of the
stellar disk strongly depends on the pitch angle of the spiral
pattern. Consequently, the disk thickness in our simulations grows
with the value of the pitch angle as seen in Figure
\ref{fig:pitchangle} and equation \ref{eq:sigmoidal}. As showed in
\citet{Jenkins1990}, where the ratio $\sigma_z/\sigma_R$ depends on
the spiral structure, here we are able to measure the shape of the
velocity ellipsoid and its dependence on the pitch angle. Because the
ratio of the velocity dispersion reaches a nearly constant value only
after a certain period of time, under the influence of the spiral
pattern (Figure \ref{fig:SzSrtime}), we computed the ratio
$\sigma_z/\sigma_R$ at the final stage in our simulations ($5\Gyr$)
for several pitch angles, for the Sc galaxy. In Figure
\ref{fig:ScSzSr50} we show how the shape of the velocity ellipsoid
would be as a function of the pitch angle. It decreases first as the
pitch angle grows but after a $40^o$ angle it fluctuates around some
constant value.

For small pitch angles ($<20^o$), the radial and vertical dispersions
are nearly the same and this relation continues after $5\Gyr$, despite
the presence of the spiral pattern. This is because even when
$\sigma_z$ and $\sigma_R$ increase with time, the grow rate is small
in both directions compared with the initial dispersions. On the other
hand, Figure \ref{fig:ScSzSr50} shows that for larger pitch angles,
not only the increment in $\sigma_z$ is considerable, as it is found
in section \ref{PitchAngle}, but the heating in the $R$-direction is
remarkably, even greater than that in the $z$-direction. This is
expected from a structure that rotates in the plane of the galaxy, but
here we have found that it has a considerable effect in both, the
radial and vertical directions, and variates sensibly with the
parameters of the spiral pattern.

\subsubsection{Transient spiral arms}
All results presented in the previous sections are obtained from
simulations that assume long-lived spiral arms. However, it is a
result from N-body simulations, that the spiral arms might rather be
transient features in general \citep{Sellwood2011,Kawata2011}.

In this section we present some experiments with transient spiral arms
now, to quantify its effect in the vertical structure of the stellar
disk and to study possible differences with our previous results. The
adjustability of the galaxy models allows us to emulate transient and
recurrent spiral arms by making them grow and disappear
periodically. The lifetime of the spiral patterns is hard to determine
in N-body simulations, and even in the cases where it can be
determined with Fourier modes analysis, it is different for every
simulation. We take two different periods for a simple experiment, one
with $100\Myr$ and other with $500\Myr$ \citep{Grand2014}.

\begin{figure}
\begin{center}
\includegraphics[width=9cm]{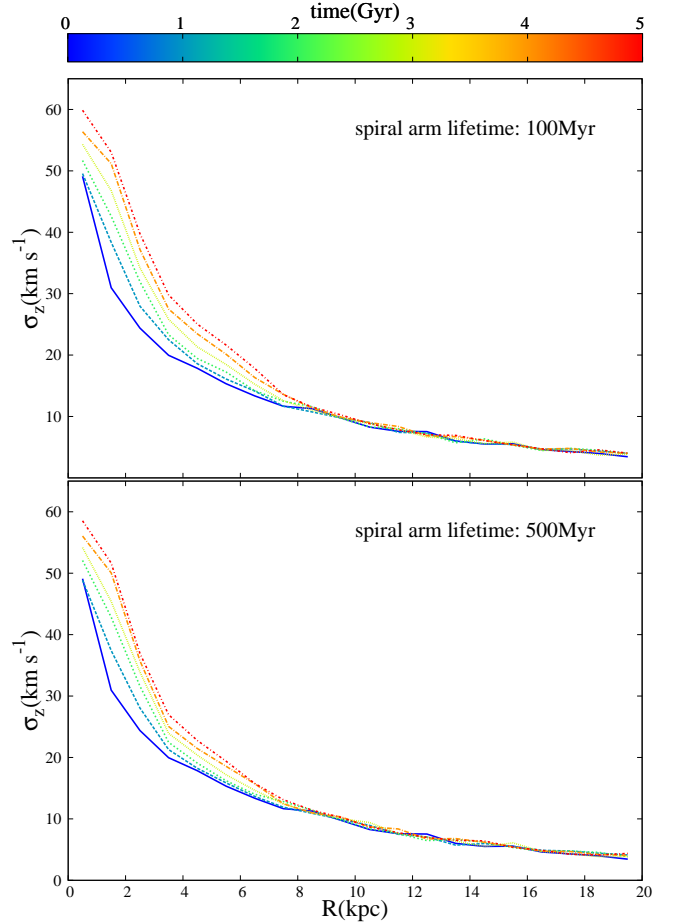}
\end{center}
\caption{Vertical velocity dispersion as a function of $R$ along
  $5\Gyr$ evolution time for transient spiral arms with lifetimes $100\Myr$
  and $500\Myr$. }
\label{fig:transient}
\end{figure}

The simulations are made with an Sc galaxy model with a spiral arms
pitch angle of $50^{\circ}$, $M_{arms}/M_{disk} = 0.05 $, and two
different lifetimes for the transient spiral arms. Figure
\ref{fig:transient} shows the vertical velocity dispersion $\sigma_z$
as a function of $R$ for the $5\Gyr$ evolution time with the transient
spiral arms. Even with this kind of spiral pattern that is not always
present in the disk, grows and disappear periodically, we can see that
$\sigma_z$ keeps increasing with time for both lifetimes used, as it
does in our previous simulations with longer lasting spiral arms.

\begin{figure}
\begin{center}
\includegraphics[width=9cm]{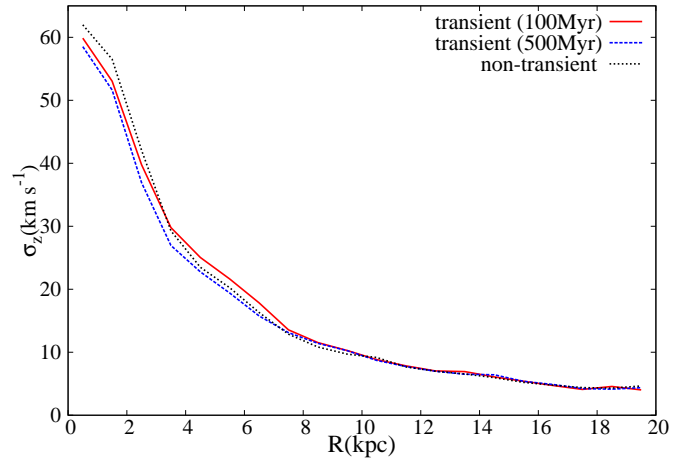}
\end{center}
\caption{A comparison of $\sigma_z$ after $5\Gyr$ for transient spiral
  arms, $100\Myr$ lifetime, $500\Myr$ lifetime, and non-transient
  spiral arms.}
\label{fig:compare}
\end{figure}

In Figure \ref{fig:compare} we compare the final vertical velocity
dispersion for both models of transient spirals against that obtained
for non-transient spiral arms. The plots show only slight deviations
between them, and not significative at all radii.

\begin{figure}
\begin{center}
\includegraphics[width=9cm]{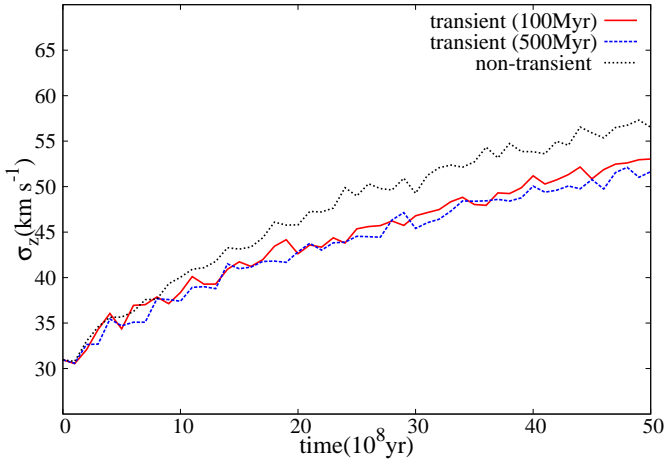}
\end{center}
\caption{Comparing the time evolution of $\sigma_z$ for transient
  spiral arms, $100\Myr$ lifetime, $500\Myr$ lifetime, and
  non-transient spiral arms.}
\label{fig:transienttime}
\end{figure}

For a more detailed comparison we pick the radial zone at which the
increase in vertical velocity dispersion is highest and plot the
temporal evolution of $\sigma_z$ for the transient and non-transient
spiral arms and show it in Figure \ref{fig:transienttime}. Some slight
differences are visible at the end of the simulation, and as this
radial zone is where $\Delta\sigma_z$ is maximum, the differences are
even smaller for the rest of the disk.

Despite the different nature of the three types of spiral arms
simulated, the induced stellar dynamical behavior is similar. This is
because the transient spiral arms, although not always present in the
disk, are recurrent and form very quickly once the previous spiral
pattern disappear as seen in N-body simulations, leaving the stars
under an almost constant influence.

\subsection{Plausible Origin of the Vertical Stellar Heating}
Based on the set of experiments presented here, we conclude that a
spiral structure can excite, considerably, velocities in the
z-direction. Furthermore, we noticed that the physical mechanism
causing the heating is different from simple resonant excitation. The
spiral pattern induces chaotic behavior not linked necessarily to
resonances, but rather to direct scattering of disk stars, which leads
to an increase of the velocity dispersion. 

\begin{figure*}
\begin{center}
\includegraphics[width=19cm]{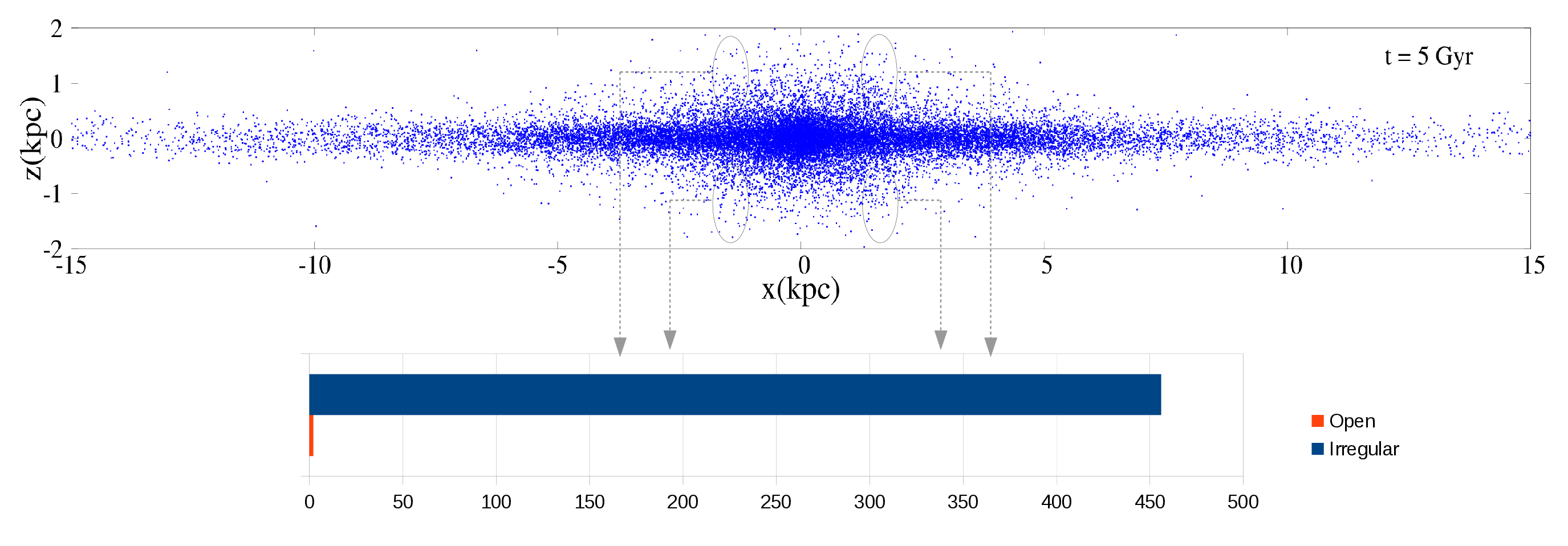}
\end{center}
\caption{Orbital classification on the disk particles that experienced a significant
increase in their velocity dispersion. The histogram shows that the orbital sample is
 dominated by irregular orbits.}
\label{fig:clasificacion}
\end{figure*}

In order to produce evidence to support this gravitational scattering
interpretation, we performed the following analysis. For an Sc galaxy
with spiral mass $M_{arms} = 0.05\ M_{disk}$ and a pitch angle of
50$^{\circ}$, we took a sample of 458 particles that at the end of the
simulation are part of the hot component (i.e. particles that
experienced a significant increase in their velocity dispersion).  By
tracing back the initial positions, we reconstruct the orbit of each
particle. To classify the orbits and distinguish them as regular or
irregular, we implemented a known spectral method by
\citet{Carpintero1998}. The method is able also to identify loop, box,
and other resonant orbits. The results are presented in Figure
\ref{fig:clasificacion}.

This analysis shows that the orbital sample is completely dominated by
irregular orbits, with no evidence of resonances from any kind. In
this way we interpret the vertical heating as caused by gravitational
scattering of stars by spiral arms, as already has been pointed out in
other studies \citep{Fujii2011}.

\subsection{The case of the Milky Way}
Probably, the hardest case to study due to the plethora of information
available, sometimes contradictory, some other model or observations
dependent, is the Milky Way Galaxy. The existence of different types
of disks, or different types or vertical structures (the old and young
thin disks, and the thick disk) has been known but not for long. What
are the mechanisms that produce the heating of stellar orbits into the
thick disk?, what is the relative importance of each one? is this
present in all galaxies? the mechanisms that affect radially the disks
are effective in the vertical component?

A lot of the finest work on mechanisms to heat dynamically the disk
has been done in the plane with steady two-dimensional potential
models studying fundamental physical phenomena such as the radial
migration. Regarding the vertical structure, this is a subject that
started several decades ago to be of interest in astrophysics. Only
recently, thanks to the calculation power of the new generation of
supercomputers, we have been able to produce more realistic and
detailed simulations with test particles in steady potentials and with
alive models with improved resolution, all to study with unprecedent
details the dynamics of the Galactic disk.

In this section, as a preliminary application to a specific
  galaxy, we have constructed a detailed density distribution, based
  model of the Milky Way Galaxy that includes spiral arms, with the
  purpose of exploring their isolated effect on the heating to the
  disk, and its participation in the process of production of the
  thick disk. We compare our results with other recent work dedicated
  to Milky-Way-like potentials \citep{Faure2014}. Additionally, the
  experiments presented in this section for the Milky Way, include one
  more experiment incorporating to the calculations the Galactic bar.
  We are not pretending this to be an extensive study of the Milky Way
  Galaxy, but rather a modest first approximation and preliminary
  results of only the vertical heating effects by spiral arms and bar
  on a detailed model of the Galaxy, in consequence it is important to
  mention, we are not including all the relevant references to this
  problem. In an ongoing work, we will provide a work fully dedicated
  to the Milky Way Galaxy in this context and the relevant
  references.

The parameters of the galactic models employed here, allow us to model
different morphological types and different spiral arms classes by
changing the pitch angle, the mass or even making them transient. In
this section, we produce observationally motivated models of the Milky
Way galaxy and study the response of the stellar disk to the spiral
arms. The parameters that describe the components that compose the
axisymmetric background potential, are constrained by recent estimates
of the galactic rotation curve. We adopt a pitch angle of
$15.5^{\circ}$ for the spiral arms, a mass $M_{arm}/M_{disk} = 0.03$,
and an initial time of $0.5\Gyr$ to increase up to their full mass the
spiral arms. For the first simulation presented for the Milky Way, we
do not include a bar in order to try to isolate the influence of the
spiral pattern in the stellar disk (and to compare with other similar
work).

\begin{deluxetable*}{lcc}
\tablecolumns{3}
\tabletypesize{\scriptsize}
\tablewidth{0pt}
\tablecaption{Parameters for the Milky Way Model}
\tablehead{Parameter &Value& Reference}
\startdata
 
 &{\it Axisymmetric Components} & \\
\hline
$R_0$& 8.5 kpc &  1\\
$\Theta_0$ &220 km s$^{-1}$ & 1\\
$M_B$&$1.41\times10^{10}$ M$_\odot$& 1\\
$M_D$&$8.56\times10^{10}$ M$_\odot$&1 \\
$M_H$&$80.02\times10^{10}$ M$_\odot$ $^{1}$& 1 \\
Disk scale-length & 2.5$\kpc$ & 2\\
b$_1$ $^2$ & 0.3873 kpc&1 \\
 a$_2$ $^2$ &5.3178 kpc &1\\
b$_2$ $^2$& 0.2500 kpc&1\\
a$_3$ $^2$ & 12 kpc&1\\
\hline
 &{\it Spiral Arms}&  \\
\hline
locus             & Logarithmic & 3\\
arms number      & 2 & 4,5\\
pitch angle   & 15.5$\deg$ & 4 \\
M$_{sp}$/M$_{D}$& 3\% &   \\
scale-length   & 2.5 $\kpc$  & disk based\\
Patter speed ($\Omega_{sp}$)  &-20 $\kmskpc$ & 6 \\
inner limit &3.3 kpc & based on ILR \\
outer limit & 12 kpc& based on CR\\
\hline
 &{\it Triaxial Bar}&  \\
 \hline
 major axis & 3.5 kpc&2, 7\\
 Scale length &1.7, 0.64, 0.44 kpc& 2\\
 Axial ratio &0.64/1.7, 0.44/1.7& \\
Mass& $1.41\times10^{10}$ M$_\odot$& \\
Pattern speed ($\Omega_B$)& 50 $\kmskpc$& 6 \\
\enddata

\tablenotetext{1} {Up to 100 kpc halo radius.} 
\tablenotetext{2}{b$_1$, a$_2$, b$_2$, and a$_3$ are scale lengths.}
\tablerefs{(1)\citealt{Allen1991}. 
           (2) \citealt{Freu98}.
           (3) ~Seigar \& James 1998; Seigar \et 2006.
           (4) ~Drimmel \et 2000.
           (5) ~Grosb\o l \et 2002; Churchwell et al. 2009; Elmegreen \& Elmegreen 2014.
           (6) ~Gerhard 2011.
           (7) ~Binney et al. 1997; Bissantz \& Gerhard 2002.}
\label{tab:parametersMW}
\end{deluxetable*}

\begin{figure}
\begin{center}
\includegraphics[width=9.2cm]{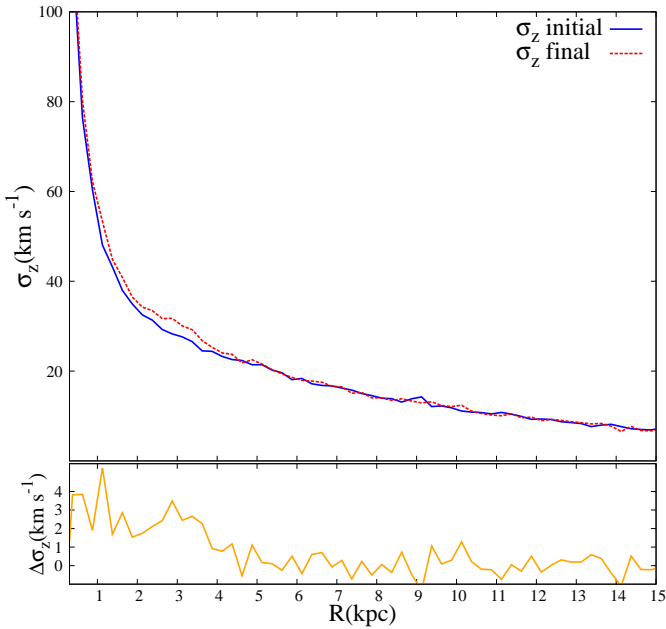}
\end{center}
\caption{Effect of the Milky-Way-like Spiral Arms in the dynamical
  heating of the disk. Top: Initial and final vertical velocity
  dispersions after $5\Gyr$ evolution. Bottom: $\Delta\sigma_z$
  (final-initial). The vertical heating is minimum over a small radial
  region.}
\label{fig:MWlike}
\end{figure}

Figure \ref{fig:MWlike} (top panel) shows the initial and final
vertical velocity dispersion after a $5\Gyr$ evolution for the
Milky-Way-like model (halo, disk, bulge and spiral arms). As seen for
the case of early type galaxies (i.e. small pitch angles and spiral
arm masses), the change in velocity dispersion is small. Indeed,
although the effect is not negligible, in the Milky-Way-like model
experiment, with a pitch angle of $15.5^{\circ}$ and a mass $M_{arm}/M_{disk} =
0.03$, the spiral arms are not capable to heat the stellar
disk. In Figure \ref{fig:MWlike} (bottom panel) we plot the difference
between $\sigma_z$ initial and final, and show that the main heating
is produced in the innermost region of the disk.

With this experiment we conclude that the vertical heating that spiral
arms are able to produce in our Milky-Way-like Galaxy model is minimum
and it is located in a small radial region. This result represents a
different one compared to that of \citet{Faure2014}, that find some
important effect on the vertical heating due to the spiral arms. To
try to account for the origin of this discrepancy, further studies
will be performed in a future work.

In a second experiment, we have also included a known mechanism that
exerts strong secular radial and vertical evolution on galactic
disks, the central bar. Although the present work is advocated to the
study of the effect of the spiral arms on stellar disks vertical
heating, we have produced a careful experiment for the Milky Way
Galaxy to explore to what extent the central bar can contribute to the
vertical heating. In an ongoing paper we study in detail the vertical
heating produced by the full galactic model of the Milky Way, here
we present some preliminary results of this study.

\begin{figure}
\begin{center}
\includegraphics[width=9.2cm]{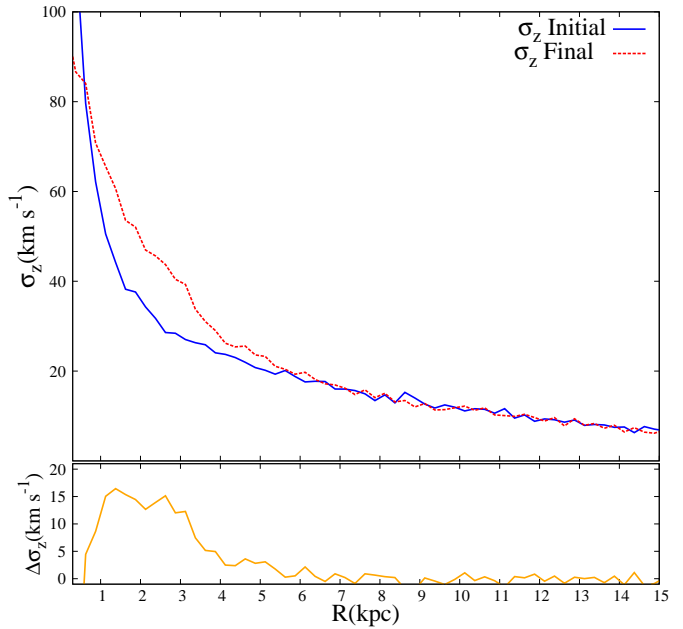}
\end{center}
\caption{Effect of the Milky Way Spiral Arms + Bar model on the
  vertical disk Heating. Top: Initial and final vertical velocity
  dispersions after $5\Gyr$ evolution. Bottom: $\Delta\sigma_z$
  (final-initial). The vertical heating is appreciable, with an
  increment up to $16 \kms$ in the inner region of the disk.}
\label{fig:MWlikeBar}
\end{figure}

We ran a simulation with the parameters representative of the Milky
Way Galaxy, given on Table \ref{tab:parametersMW}. Figure
\ref{fig:MWlikeBar} shows the initial and final vertical velocity
dispersion $\sigma_z$ across the stellar disk. Although the two
nonaxisymmetric structures are present in the simulation, by
comparing Figures \ref{fig:MWlike} and \ref{fig:MWlikeBar}, there is no
doubt that the heating seen in the second one is produced by the
galactic bar, and that the only affected region seems to be the
central part (similar results were obtained by \citep{Saha2010}).

The effect of the galactic bar on the vertical velocity dispersion is
considerably, with an increase in $\sigma_z$ up to $16 \kms$, mostly
in the inner $6\kpc$ of the disk.

\section{Conclusions} \label{conclusions}
With the use of detailed models for spiral galaxies, we produce an
extensive study of disks vertical heating. The models include an
axisymmetric component (bulge, massive halo and disk) plus a density
based three dimensional potential for the spiral arms, orbitally
tested for self-consistency. 

The main outcome of this study is that the spiral structure is capable
to excite moderate to high dispersion velocities in the $z$-direction
inducing a relation between stellar age and $\sigma_z$ in all spirals
except the earliest types. Although rather small, this mechanism works
for the Milky Way Galaxy, assuming the parameters for the spiral arms
known to this day. Consequently, by isolating the effect of the
spiral arms on the vertical velocity dispersion of the stellar disk,
we can conclude that spiral arms have the capability to contribute to
the heating mechanism that gives raise to thick disks in spiral
galaxies, from intermediate to late morphological types.

Therefore, the thickness of the stellar disk driven by the spiral arms
go from negligible to very important, depending on the characteristics
of the spiral pattern such as mass, pitch angle and angular speed
along with the morphological galaxy type. By covering a whole set of
values for this parameters in the test particle simulations we
conclude that:

\begin{itemize}

\item The relative increase in vertical velocity dispersion is a
  function of the morphological type, being smaller for early type
  galaxies (Sa type in the Hubble scheme) and larger from intermediate
  to late type galaxies. Although for the sake of clarity, we
    present our results by separating the models in approximately
    average Sa, Sb and Sc galaxy types, we are aware that strong gaps
    are in between the different models in the initial scale
    height. The results on the disk heating driven by spiral arms
    presented here, are therefore more general, i.e. the observed
    heating is considerably more significant in thinner disks, which,
    as a consequence, has implications on galaxy types, in this case,
    particularly on later types.

\item Two distinctive characteristics of the spiral arms are the mass
  and the pitch angle. By varying this parameters one at the time,
  within the permitted values, we found that they play equally
  important roles in the vertical heating of the stellar disk. As
  expected, massive arms and/or large pitch angles are responsible for
  the most prominent disk thickness found in our simulations.

\item The other dynamical parameter characteristic of the spiral arms,
  is the pattern speed $\Omega$. When varying this value, we found
  that the vertical heating induced by the spiral pattern is greater
  for slow rotating arms and decrease for those that rotate
  faster. Smaller values of $\Omega$ allow the spiral arms to interact
  more effectively with the stars, and heat more efficiently the stellar disk,
  independently of galaxy type.

\item By measuring the time evolution of $\sigma_{z}$ in our
  simulations, we found that the vertical heating induced by the spiral
  arms follows a power law $\sigma_z$ $\propto$ $t^{\alpha}$, as seen
  from observations in the solar neighbourhood and other numerical
  simulations. An interesting result is that the value of ${\alpha}$
  depends only on the pitch angle and not on the mass of the arms.

\item The ratio $\sigma_z/\sigma_R$, that defines the shape of the
  velocity ellipsoid, shows a clear trend with the morphological type
  that remains during the temporal evolution. The values of this ratio
  across the entire simulations show that the heating in the
  R-direction is remarkable, always greater that in the z-direction
  for the three galaxy types. This is expected from a structure that
  rotates in the plane of the galaxy, but here we have found that it
  has a considerably effect in both, the radial and vertical
  directions. Consequently, the spiral arms are key to determine the
  shape of the velocity ellipsoid.

\item Although in this work we cover the general properties of normal
  spiral galaxies, we took advantage of the adjustability of our model
  to represent the gravitational potential of the Milky Way
  Galaxy. Analysing the change in velocity dispersion induced by the
  nonaxisymmetric structures of the Galaxy we found that: the
  galactic spiral arms are not capable to induce an important
  thickness in the stellar disk, the increase in vertical velocity
  dispersion is small, therefore, from this study we conclude that the
  spiral arms play no role in producing a thick disk. If we add the
  galactic bar on the other hand, the vertical velocity dispersion
  increase considerably, mostly within the region covered by the
  bar. This means that for the Milky Way, the bar is an important
  heating mechanism that should be considered in calculations, but
  mostly in the inner region of the disk.

\end{itemize}

\acknowledgments

We thank Octavio Valenzuela for enlightening discussions. We thank the
anonymous referee for an excellent review that helped to improve this work.
We thank PAPIIT through grant UNAM/DGAPA IN114114. CONACyT M\'exico under
grants CB-2009-01, no. 132400, CB-2011, no. 166212. The authors
acknowledge to CGSTIC at CINVESTAV for providing HPC resources on the
Hybrid Cluster Supercomputer "Xiuhcoatl". LAMM is supported by a
CONACYT scholarship. APV acknowledges the the postdoctoral Fellowship
of DGAPA-UNAM, Mexico.

\end{document}